\documentclass{article}
\usepackage{arxiv}
\usepackage[utf8]{inputenc}
\usepackage[T1]{fontenc}
\usepackage{hyperref}
\usepackage{url}
\usepackage{booktabs,longtable}
\usepackage{amsmath,amssymb,amsthm,mathtools}
\usepackage{nicefrac}
\usepackage[expansion=false]{microtype}
\usepackage{graphicx}
\usepackage[numbers,sort&compress]{natbib}
\usepackage{xcolor}
\usepackage{array}
\usepackage{tabularx}
\usepackage{placeins}
\usepackage[font=small,labelfont=bf,skip=5pt]{caption}

\newtheorem{theorem}{Theorem}
\newtheorem{proposition}{Proposition}

\newtheorem{definition}{Definition}
\newtheorem{remark}{Remark}

\newcommand{\figfile}[1]{\makebox[\linewidth][c]{\includegraphics[width=180mm]{#1.pdf}}}

\newcommand{\Eone}{\mathrm{E1}}
\newcommand{\Etwo}{\mathrm{E2}}
\newcommand{\Ethree}{\mathrm{E3}}

\newcommand{\Efive}{\mathrm{E5}}
\newcommand{\norm}[1]{\left\lVert #1\right\rVert}

\title{A mathematical theory and governed graph architecture for composable Earth and ecosystem modeling}
\renewcommand{\shorttitle}{Theory and governed graph architecture for composable Earth system modeling}

\author{
  Yuchong Long \\
  School of Agriculture and Biology \\
  Shanghai Jiao Tong University \\
  Shanghai 200240, China \\
  \texttt{ayinmostima@sjtu.edu.cn} \\
  \And
  Wenkai Xue \\
  School of Agriculture and Biology \\
  Shanghai Jiao Tong University \\
  Shanghai 200240, China \\
  \texttt{chegov.k@sjtu.edu.cn} \\
  \And
  Bo Peng \\
  School of Agriculture and Biology \\
  Shanghai Jiao Tong University \\
  Shanghai 200240, China \\
  \texttt{cnpenpoer@gmail.com} \\
  \AND
  Ji Zheng \\
  School of Agriculture and Biology \\
  Shanghai Jiao Tong University \\
  Shanghai 200240, China \\
  \texttt{zhengjiji456@sjtu.edu.cn} \\
  \And
  Shan Yin\thanks{Corresponding author: \texttt{yinshan@sjtu.edu.cn}} \\
  School of Agriculture and Biology \\
  Shanghai Jiao Tong University \\
  Shanghai 200240, China \\
  \texttt{yinshan@sjtu.edu.cn}
}

\begin{document}
\maketitle

\begin{abstract}
Projecting coupled biosphere feedbacks under climate change requires models that couple radiation, soil water, plant physiology and atmospheric chemistry from leaves to planetary grids. Contemporary ecosystem and Earth system codes join these processes through informal conventions: interfaces inspect unit strings alone, spatial aggregation relies on conserved input integrals, and close agreement with canopy fluxes is read as confirmation of internal physiology, leaving execution drift, splitting defects, scale commutation errors, support limits and mechanistic equifinality entangled in one residual. We formulate a relational theory separating model conservation structure from process parameterizations, an exact five layer calculus of discrepancy and identifiability, and their realization in the Recomposable Ecosystem Library for Integrated Coupling (RELIC), a contract governed differentiable graph compiler. We prove eight analytical limits on inference from agreement, a five layer stepwise substitution identity holding without linearization or independence, an observational equivalence theorem over unobserved supplies, and an attribution criterion isolating mechanistic residuals. RELIC compiles four abstract scientific contracts, five tuple quantity identities and declared read projections into sealed plans with conservation checked commits and automatic differentiation on graphics processors. Across 16 eddy covariance stations spanning 192 compiled compositions free of silent ancestral drift, 11 flux towers with independent temporal evaluation, and global 0.5 degree land grids from 2003 to 2024 against satellite benchmarks, this framework separates software defects, nonlinear coarsening bias and physiological equivalence orbits for ecologists, Earth system modelers and software engineers.
\end{abstract}


\section{Global change coupling and the entanglement crisis in ecosystem modeling}
\label{sec:intro}

Warming, hydrological extremes and land use change act on the planetary system through the terrestrial biosphere, where forests alter solar radiation, evaporation, carbon storage and atmospheric composition together~\cite{Bonan2024,Friedlingstein2026,Weber2024}. The exchanges that carry these feedbacks cross disciplinary borders. A single flux of water, carbon or a reactive trace gas such as isoprene depends on the solar radiation field, on water transport through soil and plant, on the biochemistry of the chloroplast and on the oxidizing capacity of the troposphere~\cite{Guenther2012,Weber2022}. Heat and drought change stomatal conductance~\cite{Medlyn2011} and with it evaporation~\cite{Denissen2022}, carbon uptake and the deposition of ozone; the same stress alters the emission of reactive carbon~\cite{Wang2022Drought,Morfopoulos2022}, with temperature sensitivities that differ between ecosystems~\cite{Seco2022,Wang2024Arctic,DiMaria2023}, and the emission feeds back on ozone and aerosol. Predicting the response of the Earth system to global change therefore means composing radiative, hydrological, physiological and chemical processes across leaves, canopies and model grid cells, each formulated at its own scale and by its own community. Biogenic trace gas exchange, which involves all four domains at once, is a demanding test of whether that composition is sound.

Land surface and ecosystem models assemble such processes from modules written by different communities and joined by coupling frameworks~\cite{Hill2004,Valcke2013,Liu2023CCoupler3,Lawrence2019,Schneck2022}. The modules exchange arrays through interfaces that were designed for performance and portability, and the scientific meaning of an exchanged quantity is carried by convention. Three assumptions usually travel with this practice without proof. The first concerns interfaces: if the unit strings of two ports agree and the budget closes at the end of a run, the executed code is taken to represent the declared equations. The second concerns scale: if a remapping operator conserves the integral of each input, the coarse simulation is taken to approximate the fine one. The third concerns validation: if a richer physiological or chemical formulation~\cite{Knauer2023} reproduces tower fluxes or satellite columns, its parameters are taken to be confirmed by observation, although the limits of confirming open numerical models and the equifinality of calibrated ones have long been recognized~\cite{oreskes1994,beven2006}, the parameters of land models remain poorly constrained~\cite{Kennedy2025,Raoult2025}, forcing the design of ecosystem model ensembles to confront structural equifinality directly~\cite{Guan2026}.

\begin{figure}[tp]
\centering
\figfile{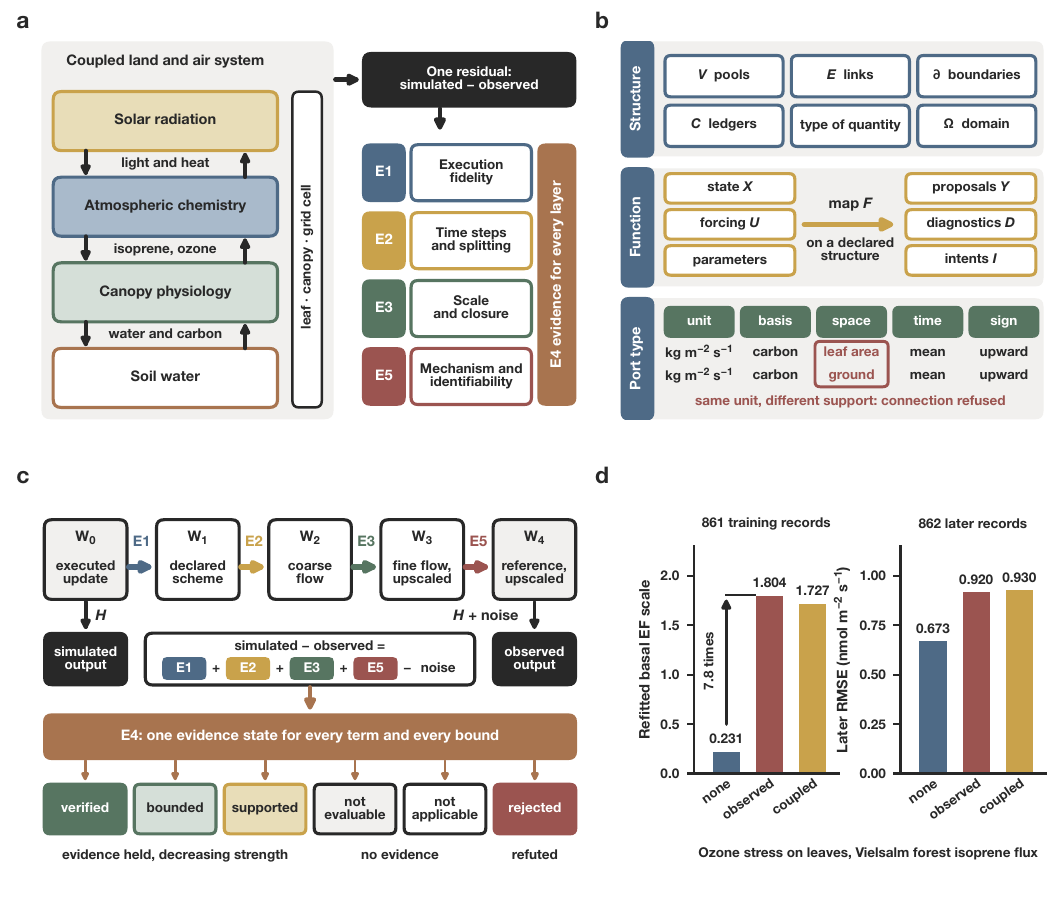}
\caption[Five layer entanglement and the contracts of model structure and process function]{\textbf{Five layer discrepancy entanglement and the contracts of model structure and process function.} \textbf{a}, Radiation, atmospheric chemistry, canopy physiology and soil water coupled from leaf to grid cell, and the four replacement layers E1, E2, E3 and E5 that accumulate in one residual between simulation and observation, each qualified by evidence on the E4 evaluation axis. \textbf{b}, The structural sextuple $(V,E,\partial,C,\tau,\Omega)$, the functional map from state, forcing and parameters to proposals, diagnostics and intents, and the five fields of the port type $\tau$; two ports with one unit string and different spatial support are refused. \textbf{c}, The replacement chain $W_0$ to $W_4$ of Theorem~\ref{thm:telescope} under a fixed comparison contract, the observation map $H$ with noise $\eta_{\mathrm{obs}}$, the exact residual identity and the six evidence states of E4. \textbf{d}, Measured compensation at the Vielsalm tower: basal emission factor (capacity) multipliers refitted on the first 861 records without ozone stress on leaves, with observed and with coupled ozone stress (0.231, 1.804 and 1.727), and the error on the 862 later records (0.673, 0.920 and 0.930~nmol~m$^{-2}$~s$^{-1}$). Panels a to c are schematics. Datasets and benchmark configurations are summarized in Table~\ref{tab:provenance}.}
\label{fig:one}
\end{figure}

These assumptions hide a common difficulty. The difference between a simulated value $y_{\mathrm{sim}}$ and an observed value $y_{\mathrm{obs}}$ is one number, yet it accumulates over five distinct layers (Fig.~\ref{fig:one}a,c): the fidelity of execution to the declared semantics (E1), time discretization and operator splitting at one scale (E2), representation and closure across spatial and temporal scales (E3), the support and evaluation status of independent evidence on an orthogonal evaluation axis (E4), and the constitutive form of process parameterizations together with the identifiability of their parameters (E5). A measured contrast shows what is at stake (Fig.~\ref{fig:one}d). At the Vielsalm forest tower, an emission law without ozone stress, fitted to the first 861 flux records, needs a basal emission factor (capacity) multiplier of 0.231. Switching on a stomatal ozone stress that no independent measurement constrains raises the refitted multiplier by a factor of 7.8, to 1.804. On the 862 later records the error grows from 0.673 to 0.920~nmol~m$^{-2}$~s$^{-1}$. A calibrated scalar has absorbed a structural change, and the more elaborate model predicts worse. Without a layered contract and a theory of identifiability, such compensation is easily read as physiological evidence, and a later calibration of another module can absorb it again.

Three research traditions have addressed parts of this problem. Theoretical ecology established the ecosystem as an open system of stocks and transfers~\cite{lindeman1942}, showed how complexity bears on stability~\cite{may1972}, and identified pattern and scale, design compromises in model building and the distortion caused by aggregating nonlinear relations as central questions~\cite{levin1992,levins1966,rastetter1992}, the last of which persists as the representation of heterogeneity below the grid cell~\cite{TorresRojas2022}; it did not supply execution semantics for large coupled codes, whose ecological complexity keeps growing~\cite{Bonan2024}. Earth system software engineering produced couplers and conservative remapping libraries of high performance~\cite{Jones1999,UllrichTaylor2015,Mahadevan2022,Taylor2024}, which govern array exchange between components while leaving the conservation topology of constitutive laws and error attribution across layers unspecified. Machine learning guided by physics and differentiable programming for the geosciences now ask for gradients through coupled process chains~\cite{Reichstein2019,Shen2023,Gelbrecht2023,Aboelyazeed2023} and hybrid weather and climate models~\cite{Schneider2023,Kochkov2024,Eyring2024Multiscale,Bracco2024}, yet handwritten scripts can break conservation and adjoint consistency at clipped boundaries, device transfers or undeclared state reads. None of the three traditions states what must be held fixed for two coupled models to be compared, how the error of a composition follows from the errors of its parts, or which differences between mechanisms a given observation design can detect at all.

We address this gap from first principles through three questions. The question of formalization asks for the minimal complete contract separating model conservation structure from process parameterizations that a composable ecosystem or Earth system model requires. Throughout this study, model structure ($\mathcal{S}$) denotes the conservation control volumes, coupling topology, standardized quantity metadata and admissible domains of a numerical model, whereas process function ($\mathcal{F}$) denotes its constitutive flux parameterizations, distinguishing modeling architecture from community ecology's classical pairing of species composition and ecosystem functioning. The question of attribution and identifiability asks how the contributions of E1 to E5 can be decomposed exactly in strongly nonlinear coupled systems, and when a residual is entitled to refute or separate mechanisms. The question of architecture asks how that mathematics can be compiled without loss into scientific software that supports conservation checked state commits at several rates, bitwise replay, tensor batching on graphics processors and automatic differentiation (Fig.~\ref{fig:one}b).

The answers are developed in order. Section~\ref{sec:theory} states the relational theory, a composable refinement of the axiomatic programme of \emph{Methods of Ecological Analysis}~\cite{Long2026MethodsEcologicalAnalysis}. Section~\ref{sec:limits} proves eight analytical limits on inference from model agreement, each with an analytical benchmark case that can be checked by hand. Section~\ref{sec:calculus} derives the five layer stepwise substitution (telescoping) decomposition, the theorem on observational equivalence and the attribution criterion. Section~\ref{sec:architecture} presents the Recomposable Ecosystem Library for Integrated Coupling (RELIC), whose compiler and runtime enforce these contracts. Section~\ref{sec:verification} verifies theory and architecture at 16 flux stations, evaluates them at 11 towers with independent temporal records and tests global 0.5 degree land grids for 2003 to 2024, and Section~\ref{sec:discussion} synthesizes the consequences for ecologists designing observation networks, Earth system modelers and scientific software engineers.

\section{Relational theory of model conservation structure, process function and quantitative composition}
\label{sec:theory}

A comparison is specified by its objects, admissible domains and metrics. Continuous scientific states are finite tuples normalized by declared positive reference scales, and the coordinatewise maximum norm is used unless another norm is stated. Discrete graph identities, modes and parameter bindings are fixed parts of the comparison contract, and gains carry the unit ratios of their input and output metrics.

\subsection{Model conservation structure, process function and quantity identity}

\begin{definition}[Ecosystem model structure]\label{def:structure}
The ecosystem model structural sextuple is
\[
\mathcal S=(V,E,\partial,C,\tau,\Omega).
\]
$V$ is the set of pools, fields, prognostic state variables and control volumes. $E$ contains the directed transfers of matter and energy, information flows and causal couplings. The operator $\partial$ fixes orientation, boundaries and the pairing of sources with sinks. $C$ encodes conservation, elemental stoichiometry and mass or water budget closure. The assignment $\tau$ gives every quantity its standardized identity (Definition~\ref{def:type}), and the joint domain $\Omega$ contains the admissible states, parameters and forcings, including positivity and thermodynamic admissibility.
\end{definition}

A schematic diagram of pools and arrows is a partial structure. A complete structural contract also locks object identities, coupling orientations, control boundaries, units, elemental bases, supports, single authoritative state write ownership and conservation pairings. To prevent terminological ambiguity with community ecology, where ecosystem structure denotes species composition or trophic architecture and ecosystem function denotes aggregate fluxes or services, $\mathcal S$ and $\mathcal F$ here denote modeling conservation architecture and constitutive process parameterizations, respectively (Table~\ref{tab:bridge}).

\begin{definition}[Ecological process parameterization function]\label{def:function}
Given a declared structure $\mathcal S$, a process parameterization (functional relationship) is the constrained map
\[
\mathcal F:D_{\mathrm{joint}}\subseteq X\times U\times\Theta\longrightarrow Y\times\mathcal D\times\mathcal I ,
\]
where $X$ is the scientific prognostic state space, $U$ the atmospheric forcing space, $\Theta$ the parameter space, $Y$ the space of fluxes or proposed state increments, $\mathcal D$ the diagnostic space and $\mathcal I$ the space of intents for external operations, which are delivered only after a scientific step has been accepted. $D_{\mathrm{joint}}$ is the admissible input domain permitted by $\mathcal S$.
\end{definition}

\begin{definition}[Five tuple quantity identity]\label{def:type}
Every state variable and coupling port carries the ordered five tuple
\[
\tau=(\mathrm{unit},\mathrm{basis},\mathrm{supp}_x,\mathrm{supp}_t,\mathrm{sign}),
\]
where $\mathrm{unit}$ names a dimension with scale and offset, $\mathrm{basis}$ the chemical accounting basis (whole molecule $\mathrm{CO_2}$ or elemental $\mathrm C$), $\mathrm{supp}_x$ the spatial support (projected ground area or leaf area on one side), $\mathrm{supp}_t$ the temporal meaning (instantaneous value, interval mean or interval total) and $\mathrm{sign}$ the positive direction across an explicit control boundary.
\end{definition}

Agreement of all five fields makes two ports directly compatible, and a registered conversion makes them castable on its declared domain. The unit string alone is a lossy projection of $\tau$, which is the content of Proposition~\ref{prop:unit}.

\subsection{Comparison contract and the five layers}

\begin{definition}[Common comparison contract]\label{def:contract}
The common comparison contract is the tuple
\[
\mathcal C=(X,U,\Theta,x_0,b,R,H,D,\norm{\cdot}_D).
\]
The external conditions $\eta=(u,\theta,b,h,\mathrm{window},\mathrm{support})$ are fixed, including boundary and sign conventions $b$ and step size $h$. The comparison domain $D=D_\eta$ is the admissible state section at fixed $\eta$, with $x_0\in D_\eta$. The spatial upscaling (aggregation or restriction) operator $R:X_f\to X_c$ maps a fine to a coarse representation, and the observation operator $H:X_c\to Y_{\mathrm{obs}}$ maps coarse states to observed quantities. A change of forcing, parameters, boundary or support defines a different comparison section and is never conflated with a change of mechanistic form.
\end{definition}

\begin{definition}[Local discrepancies E1, E2, E3 and E5 and the evaluation axis E4]\label{def:layers}
Let $\mathcal Z$ be the runtime state space, $Y_{\mathrm{num}}$ the complete numerical state space and $X$ the scientific state space, with $\alpha:\mathcal Z\to Y_{\mathrm{num}}$, $\beta:Y_{\mathrm{num}}\to X$ and $\pi=\beta\alpha$. The numerical state includes the history, stages and held inputs that the declared scheme $\mathsf S_h:Y_{\mathrm{num}}\to Y_{\mathrm{num}}$ requires.
\begin{enumerate}
\item \emph{E1, fidelity of execution.} For admissible $z$, $\varepsilon_{\mathrm{exec}}=d_X\bigl(\pi\mathcal T_{\mathrm{exec}}(z),\beta\mathsf S_h(\alpha z)\bigr)$ compares the executed update with the declared scheme.
\item \emph{E2, discretization and coupling at one scale.} With continuous flow $\Phi_{\mathcal S,\theta}$ and admissible numerical history $y$, $\varepsilon_{\mathrm{num}}=d_X\bigl(\beta\mathsf S_h(y),\Phi_{\mathcal S,\theta}(\beta y)\bigr)$. For composed operators with a valid nonnegative gain matrix $K_k$ and local defects $\eta_k$, the error vector obeys $\mathbf e_{k+1}\le K_k\mathbf e_k+\eta_k$.
\item \emph{E3, representation and closure across scales.} With fine and coarse flows $\Phi_f,\Phi_c$, downscaling (reconstruction or prolongation) operator $P$ and extensive ledgers $C_f,C_c$, three quantities are reported separately: the conservation defect $d_I=\norm{C_cRx-C_fx}$, the commutation residual $e_F=\norm{\Phi_c(Rx)-R\Phi_f(x)}_{X_c}$ and the reconstruction loss $e_R=\norm{PRx-x}_{X_f}$.
\item \emph{E5, constitutive form and identifiability.} With candidate $\Phi_\theta$ and declared reference $\Phi^\star$ on a locked common support, $\varepsilon_{\mathrm{struct}}=\norm{\Phi_\theta-\Phi^\star}_D$.
\end{enumerate}
E4 is an orthogonal independent evaluation axis. It maps equation residuals, a posteriori bounds, reference benchmarks and independent observations to six evidence states: \texttt{verified}, \texttt{bounded}, \texttt{supported}, \texttt{not\_evaluable}, \texttt{not\_applicable} and \texttt{rejected}.
\end{definition}

The order E1, E2, E3, E5 follows the prerequisites of attribution: a residual can be read as mechanism only after execution, discretization and scale have been bounded and verified. Table~\ref{tab:layers} collects the objects compared in each layer, and Table~\ref{tab:bridge} aligns RELIC formal objects and E1 to E5 with classical Earth system and ecological terminology.

\begin{table}[t]
\centering
\small
\caption[The five layer discrepancy and evidence framework]{\textbf{The five layer discrepancy and evidence framework.} Objects compared, metric, analytical benchmark case and the node of the RELIC architecture that governs each layer.}
\label{tab:layers}
\begin{tabularx}{\linewidth}{@{}l>{\raggedright\arraybackslash}X>{\raggedright\arraybackslash}X>{\raggedright\arraybackslash}X@{}}
\toprule
Layer & Compared objects and metric & Analytical benchmark case & Governing node \\
\midrule
E1 & executed update against declared scheme; $\varepsilon_{\mathrm{exec}}$ & 44 g of $\mathrm{CO_2}$ booked as carbon; donor clipped on one side creates 2 g C & port proof and sealed plan of \texttt{compile\_graph}; declared input projection \\
E2 & declared scheme against continuous flow; $\varepsilon_{\mathrm{num}}$ and gain recurrence & Lie defect $\tfrac12h^2[A,B]y$ against Strang $O(h^3)$ & rate groups and commits of \texttt{StateManager}; interval enclosures \\
E3 & coarse flow of upscaled state against upscaled fine flow; $d_I$, $e_F$, $e_R$ & $d_I=0$ with $e_F=(a-b)^2/4>0$ & \texttt{ScaleCoupling} contracts and conservative remapping \\
E4 & residuals, bounds, references and independent data; six evidence states & stock changes blind to circulation; three observation windows of one wave & verification rules, checkpoints and provenance records \\
E5 & candidate against reference on locked support; $\varepsilon_{\mathrm{struct}}$ & orbit $(V,s)\mapsto(aV,s/a)$ & parameter system, gradients and Jacobian rank analysis \\
\bottomrule
\end{tabularx}
\end{table}

\begin{table}[t]
\centering
\small
\caption[Terminology bridge between RELIC formal objects and classical Earth system and ecological frameworks]{\textbf{Terminology bridge between RELIC formal objects and classical Earth system and ecological frameworks.} Correspondence across modular coupling standards (ESMF, NUOPC, CSDMS BMI and SUMMA), classical uncertainty categories and Oreskes verification versus evaluation boundaries.}
\label{tab:bridge}
\begin{tabularx}{\linewidth}{@{}>{\raggedright\arraybackslash}p{0.25\linewidth}>{\raggedright\arraybackslash}X>{\raggedright\arraybackslash}p{0.26\linewidth}@{}}
\toprule
RELIC object or layer & Classical Earth system and ecological terminology & Reference standard or boundary \\
\midrule
Structure $\mathcal S$ vs. function $\mathcal F$ & model conservation and coupling architecture vs. process parameterizations (distinct from community structure and ecosystem functioning) & SUMMA modular separation; land surface models \\
Quantity identity $\tau$ and port contracts & standardized component interfaces, variable and metadata harmonization, CF style units and area bases & ESMF, NUOPC, CSDMS BMI, C-Coupler~\cite{Hill2004,Liu2023CCoupler3} \\
Process bridges and \texttt{FunctorBase} & modular process components or submodels with prognostic states and diagnostic fluxes & ESMF components; CLM~\cite{Lawrence2019} \\
E1 (execution fidelity) & implementation, interface harmonization and mass or water budget closure error & code verification~\cite{oreskes1994} \\
E2 (same scale numerics) & temporal discretization, operator splitting (Lie or Strang) and multirate subcycling error & numerical verification~\cite{strang1968,oreskes1994} \\
E3 (cross scale representation) & area weighted upscaling ($R$), downscaling ($P$), conservative remapping, subgrid heterogeneity and representation error & conservative remapping; mosaic and subgrid scaling~\cite{Jones1999,rastetter1992} \\
E4 (evaluation axis) & independent observational evaluation, benchmarking, observation error and a posteriori error bounding & evaluation and benchmarking~\cite{oreskes1994,Estep1995Posteriori} \\
E5 (constitutive form) & model structural uncertainty, parameter equifinality and structural versus practical identifiability & equifinality and identifiability~\cite{beven2006,Miao2011} \\
External channel $\eta$ & atmospheric forcing, initial or boundary condition and prior parametric uncertainty & classical forcing and parameter error sources \\
\bottomrule
\end{tabularx}
\end{table}

\subsection{Six physical conservation laws and structural invariants}

Six fundamental physical conservation laws and structural invariants give RELIC its name. Each scientific claim requires the applicable invariants to be verified on its declared domain.
\begin{enumerate}
\item \emph{Incidence conservation.} On the physical exchange graph of one conserved element, with pools as nodes and internal transfers as directed edges, every column of the incidence matrix $\mathbf B$ holds $-1$ at the donor and $+1$ at the recipient. The identity $\mathbf 1^\top\mathbf B=\mathbf 0$ conserves total mass whatever the constitutive flux law.
\item \emph{Thermodynamic direction of open systems.} Internal entropy production is nonnegative, and organization away from equilibrium is sustained by exchange across boundaries; port-Hamiltonian and general equation for nonequilibrium reversible and irreversible coupling forms express the constraint~\cite{Oettinger2005,MehrmannUnger2023}, also for learned dynamics~\cite{Zhang2022GFINNs}.
\item \emph{Cycle topology.} For a graph with $n$ nodes, $m$ edges and $\kappa$ connected components, the cycle space has dimension $\beta_1=m-n+\kappa$, the first Betti number.
\item \emph{Unobservable circulation.} The discrete Hodge decomposition splits edge fluxes into orthogonal gradient and circulation parts~\cite{jiang2011}. A circulation satisfies $\mathbf B\mathbf c=\mathbf 0$, so $\mathbf B(\mathbf f+\mathbf c)=\mathbf B\mathbf f$ and stock changes are blind to internal cycling.
\item \emph{Governed evaluation.} A process call is the ordered composition $\Psi_{\mathrm{gov}}=\mathrm{post}\circ\Psi_{\mathrm{ker}}\circ\mathrm{pre}$ of input checks, an arithmetic kernel without side effects and output checks. It yields state proposals, diagnostics and intents; the single authoritative state owner verifies and commits the proposals.
\item \emph{Typed quantities.} Every scientific quantity carries the identity $\tau$ of Definition~\ref{def:type}.
\end{enumerate}
In the structuralist reading, a theory is a predicate of structural constraints, and an empirical claim asserts that intended systems embed into its models within a bounded error~\cite{balzer1987}.

\subsection{Quantitative composition, feedback closure and complete state}

For a module $f$ between declared domains $A$ and $B$, let $T_f$ act on runtime states, let $S_f$ be the reference map on scientific states and let $\pi_A,\pi_B$ interpret runtime states. A quantitative contract consists of finite $L_f,\epsilon_f\ge0$ with
\[
d_B(\pi_BT_fz,S_f\pi_Az)\le\epsilon_f,
\qquad d_B(S_fx,S_fx')\le L_f\,d_A(x,x').
\]
Inserting $S_g\pi_BT_fz$ between the two composite outputs and applying the triangle inequality gives the series rule
\[
L_{g\circ f}=L_gL_f,\qquad
\epsilon_{g\circ f}=\epsilon_g+L_g\epsilon_f ,
\]
which is independent of parenthesization. This perturbation algebra~\cite{HairerLubichWanner2006} is the quantitative counterpart of the wiring of open systems~\cite{baez2017,Libkind2022,Baez2023StockFlow}. Independent parallel blocks use diagonal gain blocks and concatenated defects. An admissible feedback interconnection with error vector $\mathbf e\ge0$ and a valid inequality $\mathbf e\le K\mathbf e+\boldsymbol\epsilon$, $K\ge0$, obeys the spectral gain bound
\[
\rho(K)<1\quad\Longrightarrow\quad
\mathbf e\le(I-K)^{-1}\boldsymbol\epsilon ,
\]
by the Neumann series, in line with compositional approximation of interconnected systems~\cite{RunggerZamani2018Compositional}. Eliminating an internal block $b$ from retained block $a$ gives the Schur form $K_{\mathrm{eff}}=K_{aa}+K_{ab}(I-K_{bb})^{-1}K_{ba}$, valid when $\rho(K_{bb})<1$. For $K=\left(\begin{smallmatrix}0&0.8\\0.5&0\end{smallmatrix}\right)$ and $\boldsymbol\epsilon=(0.01,0.02)^\top$ the bound is $(13/300,1/24)^\top$ with $\rho(K)=\sqrt{0.4}$.

Along a trajectory, a residual bound~\cite{Estep1995Posteriori,BartelSchaller2025} uses a continuous reconstruction $\widetilde x$ with residual $r=\dot{\widetilde x}-f(t,\widetilde x)$. A valid directional Lipschitz constant $\mu$ on the common trajectory domain gives
\[
\norm{\widetilde x(T)-x(T)}\le e^{\mu T}E_0+
\int_0^T e^{\mu(T-s)}\norm{r(s)}\,\mathrm ds .
\]
A computed contrast $\widehat\Delta$ with valid target bounds $b_0,b_1$ has a certified sign when $|\widehat\Delta|>b_0+b_1$.

Completeness of state connects these bounds to restart and identifiability. Two admissible states with the same retained information and future outputs $a,b$ force every deterministic prediction from that information to have worst error at least $\norm{a-b}/2$. Derivatives need their own check: the family $y_\delta(\theta)=y(\theta)+\delta\sin(\theta/\delta^2)$ has value error at most $\delta$ and derivative error $\delta^{-1}\cos(\theta/\delta^2)$, so agreement of values never certifies agreement of tangents.

\section{Analytical limits on scientific inference from model agreement}
\label{sec:limits}

\begin{figure}[tp]
\centering
\figfile{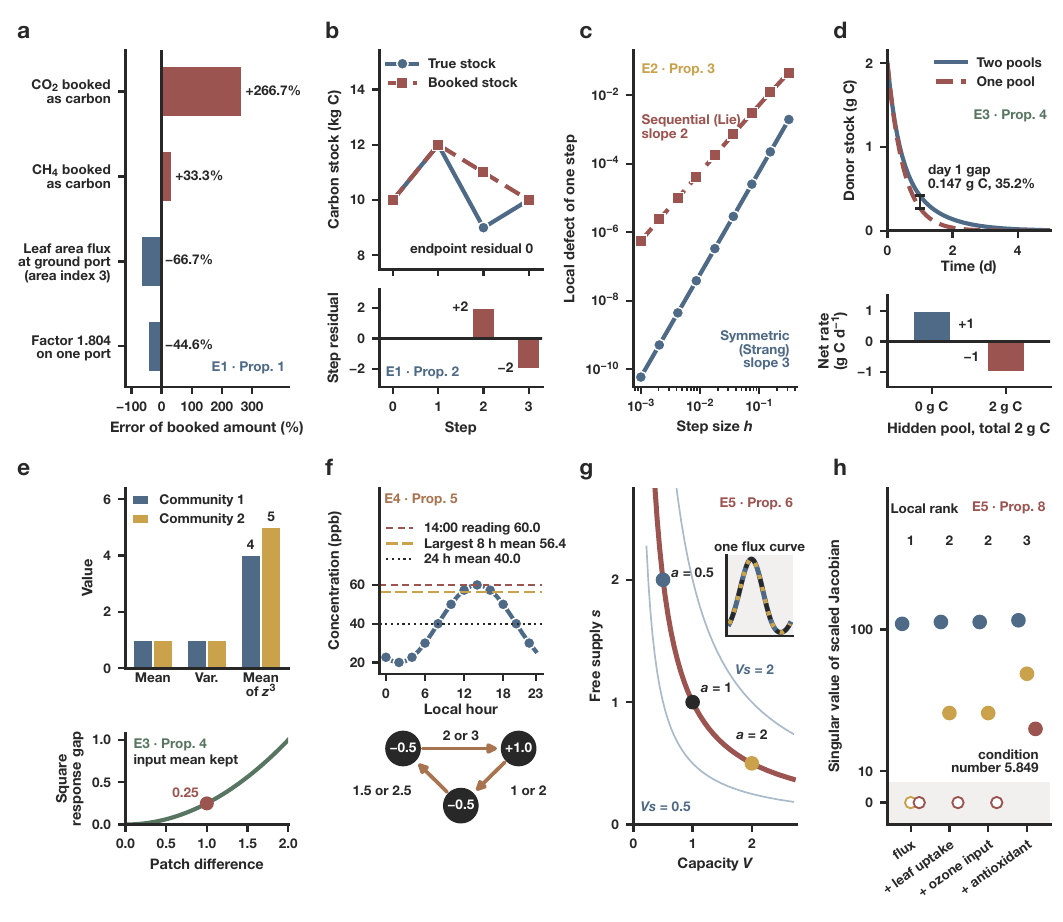}
\caption[Exact analytical benchmark cases of the eight propositions]{\textbf{Exact analytical benchmark cases of the eight propositions.} \textbf{a}, Error of a booked amount when only the unit string is checked (Proposition~\ref{prop:unit}). \textbf{b}, True and booked stocks of a ledger over three steps with zero endpoint residual and stepwise residuals of $+2$ and $-2$~kg~C (Proposition~\ref{prop:clip}). \textbf{c}, Local defect of one step of sequential (Lie) and symmetric (Strang) splitting for noncommuting generators, slopes 2 and 3 (Proposition~\ref{prop:static}). \textbf{d}, Two pools and one pool matched in initial flux under one budget of 2~g~C, and the opposite net rates of the retained total for two hidden states (Proposition~\ref{prop:three}). \textbf{e}, Two communities with equal mean and variance and cubic responses 4 and 5, and the gap $(a-b)^2/4$ of a square response under a conserved input mean. \textbf{f}, One diurnal wave under three observation windows, and a cycle of three pools whose two flux sets give identical pool rates (Proposition~\ref{prop:circ}). \textbf{g}, Hyperbolas of constant product in the plane of capacity $V$ and free supply $s$; the three marked points give one flux curve (Proposition~\ref{prop:quant}). \textbf{h}, Singular values of the scaled emission Jacobian at Vielsalm (1,723 records) for four observation sets: flux alone, with modeled leaf uptake, with prescribed ozone as an input and with an added antioxidant diagnostic; the local rank is 1, 2, 2 and 3, and values below the rank tolerance are drawn as open points (Proposition~\ref{prop:rank}). Each panel displays its layer and proposition tag from E1 through E5. Panels a to g are analytic.}
\label{fig:two}
\end{figure}

The eight propositions below are lines of defense along E1 to E5. Each marks a place where a familiar shortcut returns a false positive or a false negative, and each has an analytical benchmark case that a reader can verify by hand in under a minute (Fig.~\ref{fig:two}).

\begin{proposition}[A unit string loses meaning irreversibly; E1]\label{prop:unit}
Let $u:P\to U$ map each five tuple port record to its unit string. If two distinct records share one string, the noninjective map $u$ admits no recovery map $r:U\to P$ with $r\circ u=\mathrm{id}_P$. Equality of unit strings is one necessary field of a legal direct connection and never a sufficient one.
\end{proposition}

\emph{Analytical benchmark case.} With molar masses of 44, 16 and 12~g~mol$^{-1}$, booking 44~g of $\mathrm{CO_2}$ in a port that expects grams of carbon overstates carbon by $44/12-1=266.7\%$; the same error with $\mathrm{CH_4}$ gives $33.3\%$. A canopy with leaf area index 3 that passes a flux per leaf area to a port per ground area loses $66.7\%$ of the transfer. A calibrated factor $k=1.804$ bound to one port only leaves the reading port with the fraction $1/k$, so $44.6\%$ of the declared flux is missing (Fig.~\ref{fig:two}a).

\begin{proposition}[Clipping on one side creates mass, and endpoint budgets hide stepwise violations; E1]\label{prop:clip}
For two nonnegative closed stores with a common forward transfer $q\ge0$, both updated stores $A-q$ and $B+q$ are nonnegative if and only if $q\le A$. An overdrawn step clipped at the donor alone, $\widetilde A=\max(A-q,0)$ with $B'=B+q$, creates the mass $q-A>0$. Stepwise residuals of opposite sign cancel in the endpoint budget.
\end{proposition}

\emph{Analytical benchmark case.} With $A=10$~g~C and a proposed transfer of 12~g~C, clipping the donor at zero while the recipient receives 12 creates 2~g~C. For a true stock $(10,12,9,10)$~kg~C booked as $(10,12,11,10)$, the stepwise residuals are $(0,+2,-2)$~kg~C and the endpoint residual is 0; only the stepwise norm of 4~kg~C detects the violation (Fig.~\ref{fig:two}b).

\begin{proposition}[Static compatibility and dynamic invariance are independent, and the physical cycle graph differs from the acyclic execution graph; E1 and E2]\label{prop:static}
Let $T(p,q)\in\{\mathrm{Exact},\mathrm{Castable},\mathrm{Rejected}\}$ classify a port pair and let $f:D\to C$ be a process operator with admissible set $A\subseteq C$. Static compatibility $T(p,q)\ne\mathrm{Rejected}$ and dynamic image inclusion $f(D)\subseteq A$ are logically independent in both directions. For linear generators $A,B$ the defect of Lie splitting over one step is $\bigl(e^{hA}e^{hB}-e^{h(A+B)}\bigr)y=\tfrac12h^2[A,B]y+O(h^3)$, Strang splitting has local defect $O(h^3)$, and both coincide with the joint flow when $[A,B]=0$~\cite{strang1968,Blanes2024}.
\end{proposition}

\emph{Analytical benchmark case.} All four combinations are realizable: exact ports with a candidate stock of $-2$~kg~C pass assembly and fail at commit, and a rejected port pair can carry values that happen to stay admissible. An exchange between plant and soil with fluxes in both directions has a cycle in its physical graph, while one solver step unfolds it into an acyclic order of reads, evaluations and commits. Which process reads which time level is fixed by that order and by the declared links, so an omitted link shifts a time level without violating any budget. For two noncommuting rotation and decay generators the local defects of Lie and Strang splitting follow slopes 2 and 3 exactly (Fig.~\ref{fig:two}c).

\begin{proposition}[Conservation, commutation and reconstruction are pairwise independent; E3]\label{prop:three}
The conservation defect $d_I$, the commutation residual $e_F$ and the reconstruction loss $e_R$ of Definition~\ref{def:layers} impose distinct conditions, and the vanishing of any one implies the vanishing of no other over the declared class of upscaling (restriction), downscaling (reconstruction), ledger and evolution maps. In particular, for two patches of equal area with inputs $a,b$ and response $f(x)=x^2$,
\[
\frac{a^2+b^2}{2}-\left(\frac{a+b}{2}\right)^2=\frac{(a-b)^2}{4}\ge 0,
\]
with equality if and only if $a=b$, although averaging conserves the input integral exactly~\cite{Jensen1906}.
\end{proposition}

\emph{Analytical benchmark case.} Two communities with trait values $\{0,2\}$ (weights $\tfrac12,\tfrac12$) and $\{0,1,3\}$ (weights $\tfrac13,\tfrac12,\tfrac16$) share mean 1 and variance 1, and the cubic response $g(z)=z^3$ returns 4 and 5 (Fig.~\ref{fig:two}e). Two donors of 1~g~C decaying at 1 and 3~d$^{-1}$ follow $e^{-t}+e^{-3t}$, while one aggregated donor of 2~g~C at a rate matched in initial flux follows $2e^{-2t}$; both share the total of 2~g~C and the initial flux of 4~g~C~d$^{-1}$, yet at day one they differ by 0.147~g~C, or 35.2\% of the fine value (Fig.~\ref{fig:two}d). Eliminating a fast pool creates memory: for $x=C+D$ with $\dot x=I-kC$ and $\dot C=ax-(a+b+k)C$, at $x=2$~g~C, $I=1$~g~C~d$^{-1}$ and unit rates, the hidden states $C=0$ and $C=2$~g~C give $\dot x=+1$ and $-1$~g~C~d$^{-1}$, and the exact reduced equation $\dot x(t)=I(t)-ke^{-\lambda t}C(0)-ka\int_0^t e^{-\lambda(t-s)}x(s)\,\mathrm ds$ with $\lambda=a+b+k$ carries a Mori--Zwanzig memory kernel~\cite{chorin2009,Gupta2025}. Scalar controls complete the six directions: $R(x)=2x$, $P(c)=c/2$ with identity flows and ledgers gives $e_F=e_R=0$ and $d_I=1$ at $x=1$; identity maps with $\Phi_c(x)=x+1$ give $d_I=e_R=0$ and $e_F=1$; and a linear response on unequal patches gives $e_F=0$ with $e_R=\lvert a-b\rvert/2>0$.

\begin{proposition}[Stocks are blind to circulation, and observation windows cut different sections; E4]\label{prop:circ}
If $\mathbf B\mathbf c=\mathbf 0$, then $\mathbf B(\mathbf f+\mathbf c)=\mathbf B\mathbf f$ for every flux vector $\mathbf f$, so inventories and their rates cannot separate fluxes that differ by an admissible circulation. Distinct temporal observation operators return different summaries of one trajectory.
\end{proposition}

\emph{Analytical benchmark case.} In the cycle of three pools ($\beta_1=1$), edge fluxes $(2,1,1.5)$ and $(3,2,2.5)$ differ by a unit circulation and give the same net derivatives. For the diurnal wave $c(h)=40+20\cos\bigl(2\pi(h-14)/24\bigr)$~ppb sampled hourly, the 24~h mean is 40.0~ppb, the largest 8~h running mean 56.445~ppb and the reading at 14:00 60.0~ppb, a spread of 50\% (Fig.~\ref{fig:two}f).

\begin{proposition}[Quantifiers over sites do not commute, and a free multiplicative supply is unidentifiable; E5]\label{prop:quant}
For sites $S$ and parameters $P$, $\forall s\,\exists p\,R(s,p)$ does not imply $\exists p\,\forall s\,R(s,p)$. For a law $y(t)=s(t)\,V\,g(C(t))$ with $V>0$, $V\,g(C(t))>0$ and a free unknown supply $s(t)>0$, the transformation $(V,s)\mapsto(aV,s/a)$ with $a>0$ leaves the whole output trajectory unchanged, so flux data alone place the capacity on a hyperbola (Fig.~\ref{fig:two}g).
\end{proposition}

\emph{Analytical benchmark case.} With two sites, two parameter values and $R$ true only for $(s_1,1)$ and $(s_2,2)$, every site has an admissible value and no value serves both, so a calibration that succeeds site by site says nothing about one transferable parameter. For the orbit, scale factors $a=0.5$, 1 and 2 applied to $(V,s)$ give one and the same flux curve at every time; the measured counterpart is the emission factor (capacity) refit of Fig.~\ref{fig:one}d and the basal emission factor range of Section~\ref{sec:towers}.

\begin{proposition}[Nonlinear cascades break pointwise additivity, and opposing layers cancel at the endpoint; E1 to E5]\label{prop:cascade}
For a composition with nonlinear downstream stages, the terminal discrepancy equals the sum of the propagated layer differences of Theorem~\ref{thm:telescope} and in general differs from the downstream image of the local differences. For scalar outputs $\bigl(\sum_kB_k\bigr)^2=\sum_kB_k^2+2\sum_{i<j}B_iB_j$, so squared errors add only when the cross terms vanish.
\end{proposition}

\emph{Analytical benchmark case.} With $\Psi_1(x)=2x$, $\Psi_1^\star(x)=x$, $\Psi_2=\Psi_2^\star=x^2$ and $x=1$, the terminal difference is $4-1=3$ while the squared local difference is 1. Two layer terms of 1 have a squared total of 4 and a sum of squares of 2. When an interface correction and a scale correction have opposite signs, close agreement at the endpoint says nothing about the size of either.

\begin{proposition}[Observation design sets the local rank of the Jacobian; E5]\label{prop:rank}
Let the observation map $\mathcal G:\Theta\subset\mathbb R^p\to\mathbb R^m$ be continuously differentiable on an open set. A differentiable curve along which the prediction is constant has its tangent in the null space of the Jacobian, and full column rank at an interior point implies injectivity on a ball around it. Full rank is sufficient and not necessary. Added observables raise the rank only if they project onto directions that the existing ones leave in the null space.
\end{proposition}

\emph{Analytical benchmark case.} The orbit $(e^sc,e^{-s}a)$ of $\mathcal G(c,a)=ca$ has tangent $(c,-a)$ and Jacobian $J=(a,c)$, so $J\theta'(0)=0$. The measured counterpart at Vielsalm, where the rank rises from 1 to 3 as observables are added, is shown in Fig.~\ref{fig:two}h and discussed in Section~\ref{sec:verification}.

\section{Five layer discrepancy calculus and observational identifiability}
\label{sec:calculus}

A residual $y_{\mathrm{sim}}-y_{\mathrm{obs}}$ combines the consequences of execution, numerical approximation, representation and model choice. A sequence of matched replacements identifies the contribution of each along a declared path, observational equivalence fixes which parameter distinctions the measurement design supports, and a quantitative exclusion of a candidate then depends on bounds for the other layers.

\paragraph{Notation.} State and output spaces are real normed spaces of finite dimension, $\norm{\cdot}$ is the declared norm of its argument's space, and $\norm{F-G}_D:=\sup_{x\in D}\norm{F(x)-G(x)}$. A candidate $M$ has fine flow $\Phi_f^{M}$ and coarse flow $\Phi_c^{M}$; for a parametric family $M=\theta$.

\subsection{The replacement chain}

Fix a comparison contract $\mathcal C$ (Definition~\ref{def:contract}), a fine state $x\in D_f$ with $Rx\in D_c$, and a runtime state $z_0$ with $\pi z_0=Rx$. Let $H$ be $L_H$ Lipschitz on a set containing all states below and $R$ be $L_R$ Lipschitz on a set containing $\Phi_f^{M}(x)$ and $\Phi_f^\star(x)$. Replace one object at a time, from the executed update to the fine reference:
\begin{align*}
W_0&=\pi\,\mathcal T_{\mathrm{exec}}(z_0) &&\text{executed coarse update,}\\
W_1&=\beta\mathsf S^{\mathrm{decl}}_{h,c}(\alpha z_0) &&\text{declared numerical update projected to the coarse scientific state,}\\
W_2&=\Phi_c^{M}(Rx) &&\text{coarse continuous flow of the candidate,}\\
W_3&=R\,\Phi_f^{M}(x) &&\text{fine flow of the candidate, then upscaled,}\\
W_4&=R\,\Phi_f^\star(x) &&\text{fine reference process, then upscaled.}
\end{align*}
The simulation reports $y_{\mathrm{sim}}=H(W_0)$ and the observation is $y_{\mathrm{obs}}=H(W_4)+\eta_{\mathrm{obs}}$ with measurement error $\eta_{\mathrm{obs}}$. The layer terms are
\[
B_{\Eone}=H(W_0)-H(W_1),\qquad B_{\Etwo}=H(W_1)-H(W_2),
\]
\[
B_{\Ethree}=H(W_2)-H(W_3),\qquad B_{\Efive}=H(W_3)-H(W_4),
\]
and the local discrepancies on this chain are $\varepsilon_{\Eone}=\norm{W_0-W_1}$, $\varepsilon_{\Etwo}=\norm{W_1-W_2}$, $\varepsilon_{\Ethree}=\norm{W_2-W_3}$ and $\varepsilon_{\Efive}=\norm{\Phi_f^{M}-\Phi_f^\star}_{D_f}$. Here $\varepsilon_{\Ethree}$ is the commutation residual $e_F$, and E4 qualifies the evidence for every comparison and bound.

\subsection{Exact five layer decomposition}

\begin{theorem}[Five layer stepwise substitution decomposition]\label{thm:telescope}
Under the setting above:
\begin{enumerate}
\item \emph{Identity.} For every admissible $x$ and $z_0$ with $\pi z_0=Rx$,
\[
y_{\mathrm{sim}}-y_{\mathrm{obs}}=B_{\Eone}+B_{\Etwo}+B_{\Ethree}+B_{\Efive}-\eta_{\mathrm{obs}}.
\]
No expansion, no linearity of any map and no statistical independence of the terms is assumed.
\item \emph{Bound.} $\norm{B_{\Eone}}\le L_H\varepsilon_{\Eone}$, $\norm{B_{\Etwo}}\le L_H\varepsilon_{\Etwo}$, $\norm{B_{\Ethree}}\le L_H\varepsilon_{\Ethree}$ and $\norm{B_{\Efive}}\le L_HL_R\varepsilon_{\Efive}$, hence
\[
\norm{y_{\mathrm{sim}}-y_{\mathrm{obs}}}\le L_H\bigl(\varepsilon_{\Eone}+\varepsilon_{\Etwo}+\varepsilon_{\Ethree}+L_R\,\varepsilon_{\Efive}\bigr)+\norm{\eta_{\mathrm{obs}}}.
\]
\item \emph{Cross terms.} For scalar outputs, with $(C_1,\ldots,C_5)=(B_{\Eone},B_{\Etwo},B_{\Ethree},B_{\Efive},-\eta_{\mathrm{obs}})$, the squared residual is $\sum_i C_i^2+2\sum_{i<j}C_iC_j$; for an inner product norm the analogous identity uses $\langle C_i,C_j\rangle$.
\end{enumerate}
\end{theorem}

This stepwise substitution (telescoping) identity holds for threshold responses, photochemical switches and clipped stress factors, because the chain uses adjacent cancellation alone (Fig.~\ref{fig:three}a,b; Appendix~\ref{app:proofs}). Each term is also measurable by a matched pair of computations: the executed update against a replay of the declared scheme for E1, the declared scheme against a refined or enclosed solution for E2, the coarse run against the upscaled fine run for E3, and the candidate against the reference on the fine grid for E5. The decomposition is therefore an experimental protocol as much as an identity, and the bound of part 2 states how local discrepancies add up through the gains of observation and spatial upscaling.

\begin{remark}[What a single residual leaves open]\label{rem:inverse}
Errors in different layers can cancel, so a small residual leaves each $\varepsilon_k$ unconstrained. A perturbation of one layer with the others held fixed isolates that layer's observed contribution; quantitative control of its local discrepancy further needs an inverse stability condition for the downstream composite. For example, $H(u,v)=u$ maps every perturbation $(0,m)$ to zero while its norm is $\lvert m\rvert$.
\end{remark}

\begin{remark}[Completeness inside a declared pipeline]\label{rem:complete}
The ledger covers deviations of the maps in the declared pipeline when that pipeline is finite and ordered along an acyclic dependency graph, every map has an ideal counterpart, domains are closed, the bridging terms are classified exhaustively and uniquely, lateral channels are registered, interactions are controlled with a declared remainder, the target metric is well defined and every term has an evidence state. Audits of scope, interface and conservation, together with external observations, can expose processes omitted from the declaration.
\end{remark}

\begin{remark}[Order of replacement]\label{rem:order}
Another order of the four replacements changes the individual terms while the sum stays fixed by $H(W_0)$ and $H(W_4)$. For $f(a,b)=a^2b$ taken from $(2,2)$ to $(1,1)$, replacing $a$ first gives terms 6 and 1, and replacing $b$ first gives 4 and 3 (Fig.~\ref{fig:three}f). A reported ledger therefore states its order.
\end{remark}

\subsection{What observations can separate}

Structural identifiability asks which parameters an observation map can distinguish at all~\cite{BellmanAstrom1970,Miao2011,Dong2023,ReyBarreiro2023}, and equifinality is its empirical face~\cite{beven2006}. For a parametric family write $\mathcal H_\theta=H\circ R\circ\Phi_{f,\theta}$ on the declared support $D_f$; a finite design $x_1,\dots,x_N\in D_f$ with $m$ observed coordinates defines $\mathcal G(\theta)=\bigl(\mathcal H_\theta(x_n)\bigr)_{n=1}^N\in\mathbb R^m$.

\begin{theorem}[Observational equivalence and identifiability]\label{thm:equivalence}
\leavevmode\par
\begin{enumerate}
\item \emph{Equivalence classes.} The relation $\theta\sim_{\mathcal M}\theta'$ if and only if $\mathcal H_\theta=\mathcal H_{\theta'}$ on $D_f$ is an equivalence relation whose classes partition $\Theta$. All classes are singletons if and only if $\theta\mapsto\mathcal H_\theta$ is injective, that is, the parameters are globally structurally identifiable with $D_f$ as the design. A finite design gives the coarser relation $\mathcal G(\theta)=\mathcal G(\theta')$, whose classes can only be larger.
\item \emph{Local rank.} Let $\Theta\subset\mathbb R^p$ be open and $\mathcal G$ continuously differentiable with Jacobian $J$. A differentiable curve along which $\mathcal G$ is constant has its tangent in the null space of $J$. If $\operatorname{rank}J(\theta_0)=p$ at an interior point, $\mathcal G$ is injective on a ball around $\theta_0$. Full column rank is sufficient and not necessary: $\theta\mapsto\theta^3$ is injective with zero derivative at 0.
\item \emph{Free multiplicative supply.} Let the observed flux be $y(t)=s(t)\,V\,g(C(t);\vartheta)$ with known forcing $C$, capacity $V>0$, $V\,g(C(t);\vartheta)>0$ on the window and an unknown supply function $s(t)>0$ without further constraint. For every $a>0$ the pair $(aV,s/a)$ gives the same output trajectory, so the class of $(V,s)$ contains the orbit $\{(aV,s/a):a>0\}$. If instead $s$ is fixed independently of $V$, $\vartheta$ is known, $s(t_1)\ne0$ and $g(C(t_1);\vartheta)\ne0$, then $V=y(t_1)/\bigl(s(t_1)\,g(C(t_1);\vartheta)\bigr)$ is unique.
\end{enumerate}
\end{theorem}

The three parts answer different questions. Part 1 says that identifiability is a property of the pair of model and design, and that a smaller design can only merge classes. Part 2 gives a computable local test, and part 3 a global obstruction that no local test can lift. No number of flux records separates capacity from an unobserved supply inside the orbit of part 3, and one independent anchor of the supply removes the orbit (Fig.~\ref{fig:three}e). A Jacobian that is rank deficient at one point shows only that local identifiability is not certified there; two parameters share a class only when $\mathcal H_\theta=\mathcal H_{\theta'}$ on $D_f$ is shown, for example by an explicit invariant orbit.

\subsection{When a residual speaks about mechanism}

The next statement is derived here from Theorem~\ref{thm:telescope} and the reverse triangle inequality. Let $b_1,b_2,b_3\ge0$ be upper bounds on $\norm{B_{\Eone}},\norm{B_{\Etwo}},\norm{B_{\Ethree}}$ certified on the E4 axis, and let $\bar\eta\ge0$ bound $\norm{\eta_{\mathrm{obs}}}$. The nonmechanistic envelope of a candidate $M$ is
\[
\Gamma^{M}:=b_1+b_2+b_3+\bar\eta .
\]
For two candidates the output separation is $\Delta^{M,M'}=\norm{y^{M}_{\mathrm{sim}}-y^{M'}_{\mathrm{sim}}}$; writing $\mathrm{SNR}_{\Efive}$ for the mechanistic signal to noise ratio when the envelope sum is positive, the discrimination criterion is
\[
\Delta^{M,M'}>\Gamma^{M}+\Gamma^{M'},\qquad \mathrm{SNR}_{\Efive}=\frac{\Delta^{M,M'}}{\Gamma^{M}+\Gamma^{M'}}>1 .
\]

\begin{proposition}[Attribution to mechanism]\label{prop:attribution}
Assume (A1) to (A3) of Appendix~\ref{app:proofs} for a candidate $M$ and write $r^{M}=y^{M}_{\mathrm{sim}}-y_{\mathrm{obs}}$.
\begin{enumerate}
\item \emph{Lower bound.} If $\norm{r^{M}}>\Gamma^{M}$, then
\[
\norm{\Phi_f^{M}(x)-\Phi_f^\star(x)}\ \ge\ \frac{\norm{r^{M}}-\Gamma^{M}}{L_HL_R}\ >\ 0,
\]
and therefore $\varepsilon_{\Efive}>0$ for $M$. If $\norm{r^{M}}\le\Gamma^{M}$ the residual gives no information on $\varepsilon_{\Efive}$.
\item \emph{Discrimination.} Under (A4), if $\Delta^{M,M'}>\Gamma^{M}+\Gamma^{M'}$, then part 1 applies to at least one of $M$ and $M'$. If (A5) also holds for $M$, part 1 applies to $M'$.
\item \emph{No separation inside a class.} If $M=\theta$ and $M'=\theta'$ with $\theta\sim_{\mathcal M}\theta'$, evaluated under (A4), then their E5 layer terms coincide and $\Delta^{\theta,\theta'}\le\Gamma^{\theta}+\Gamma^{\theta'}$, so the criterion of part 2 cannot hold.
\item \emph{Transfer to equivalent parameters.} Let part 1 apply to $\theta$, and let $\theta'$ satisfy (A1) and (A2) with gains $L'_H,L'_R>0$, the same $H$, $R$, $x$ and $\Phi_f^\star$, and $\mathcal H_{\theta'}(x)=\mathcal H_\theta(x)$. Then
\[
\norm{\Phi_{f,\theta'}(x)-\Phi_f^\star(x)}\ \ge\ \frac{\norm{r^{\theta}}-\Gamma^{\theta}}{L'_HL'_R}\ >\ 0 .
\]
\end{enumerate}
\end{proposition}

\begin{figure}[tp]
\centering
\figfile{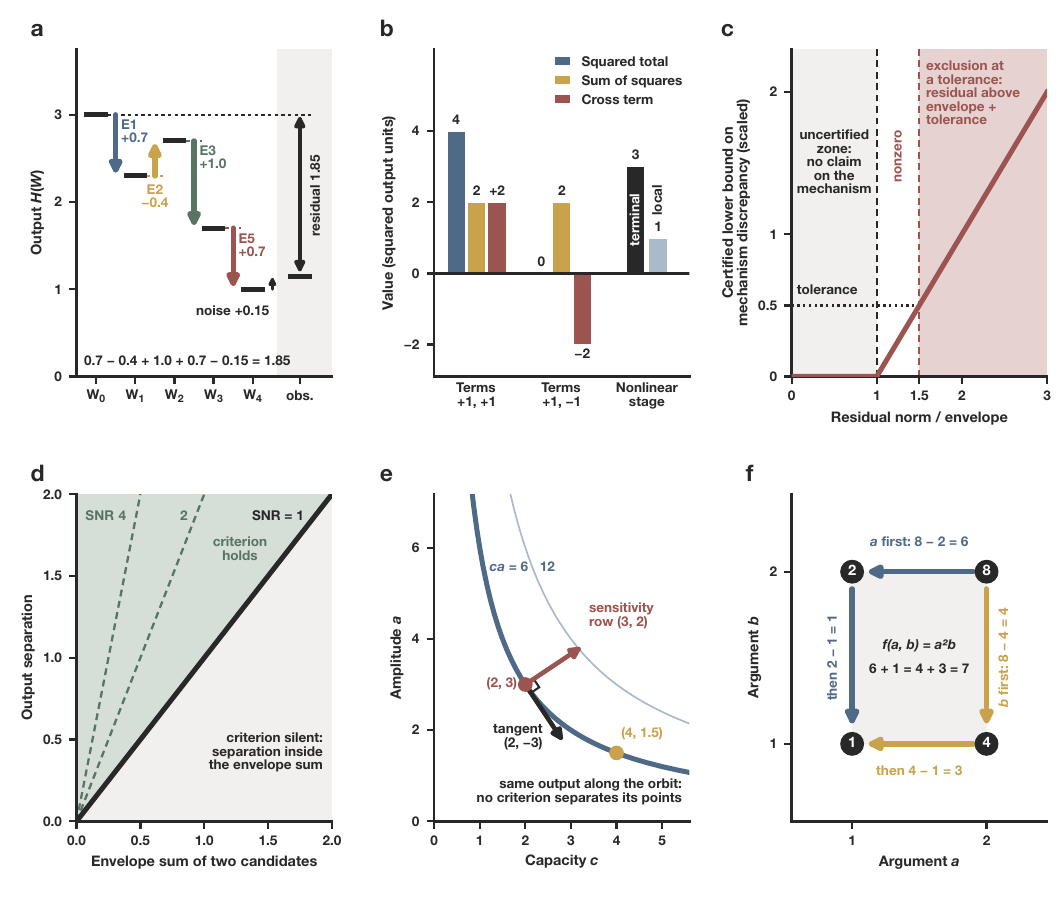}
\caption[Geometry of the five layer stepwise substitution calculus and mechanistic attribution]{\textbf{Geometry of the five layer stepwise substitution calculus and mechanistic attribution.} \textbf{a}, Replacement chain $H(W_0)$ to $H(W_4)$ in a scalar output space with constructed values: the four layer terms and the observation noise sum exactly to the residual. \textbf{b}, Cross terms: two terms of equal sign give a squared total of 4 against a sum of squares of 2, two terms of opposite sign cancel, and a nonlinear stage turns a squared local difference of 1 into a terminal difference of 3 (Proposition~\ref{prop:cascade}). \textbf{c}, Certified lower bound on the mechanistic discrepancy against the residual norm in units of the envelope $\Gamma^M$ for a unit gain product, with the shifted threshold for a tolerance $\delta_5$ of 0.5 (Proposition~\ref{prop:attribution}, part 1). \textbf{d}, Region of the plane of envelope sum and output separation where the discrimination criterion holds (part 2). \textbf{e}, The orbit $ca=6$ through $(2,3)$ and $(4,1.5)$ with the sensitivity row orthogonal to its tangent (Theorem~\ref{thm:equivalence}). \textbf{f}, Order of replacement for $f(a,b)=a^2b$ from $(2,2)$ to $(1,1)$: terms 6 and 1 or 4 and 3, both summing to 7 (Remark~\ref{rem:order}). All panels are analytic.}
\label{fig:three}
\end{figure}

\paragraph{Scientific tolerance and model families.} For a prespecified acceptable discrepancy $\delta_5\ge0$ on the fine state space, the sufficient exclusion condition $\norm{r^M}>\Gamma^M+L_HL_R\delta_5$ guarantees $\norm{\Phi_f^M(x)-\Phi_f^\star(x)}>\delta_5$. Excluding an entire family requires this for every admitted member, for example through a certified lower bound on
\[
\inf_{\theta\in\Theta_{\mathrm{adm}}}
\frac{\norm{r^\theta}-\Gamma^\theta}{L_H^\theta L_R^\theta}
\quad\text{that exceeds }\delta_5 .
\]
Part 4 transfers a nonzero discrepancy at $\delta_5=0$; at a positive tolerance each member needs its own bound or a common gain bound must cover the class. All four parts are sufficient conditions (Fig.~\ref{fig:three}c,d). The envelope is assembled from evidence of different kinds: bitwise replay or a governed comparison bounds $b_1$, a residual or interval enclosure bounds $b_2$, a matched fine run bounds $b_3$ and the instrument specification bounds $\bar\eta$, each with its evidence state on the E4 axis. The ratio $\mathrm{SNR}_{\Efive}$ then compares what the candidates disagree about with what the computation and the measurement cannot resolve. A residual below the envelope has no standing to indict the mechanism, and by part 3 the criterion separates two parameter values only when they lie in different observational classes, which added observables can create.

\section{Contract governance and differentiable graph compilation in RELIC}
\label{sec:architecture}

Coupled land and Earth system codes often degrade into a dictionary of global variables with an implicit call order, in which any routine may read any field and the coupler governs array exchange between components~\cite{Hill2004,Valcke2013,Liu2023CCoupler3}. RELIC takes the opposite position. Every mathematical object of Sections~\ref{sec:theory} to~\ref{sec:calculus} has exactly one software owner, and a scientific process can reach the state, the parameters and the device only through that owner (Fig.~\ref{fig:four}a, Table~\ref{tab:owners}).

\subsection{Design principles and package ownership}

The library is divided into packages with one responsibility each: variable contracts, the core abstractions with simulator and state management, the parameter system, the orchestration of graph declaration, compilation and execution, coupling and scale transformation, the scientific processes, solvers and operators, validation, calibration with sensitivity analysis, and extension loading. Their names in the code follow these responsibilities, from \texttt{relic.contracts} and \texttt{relic.core} to \texttt{relic.orchestration} and \texttt{relic.extensions}. Three rules hold throughout: one scheduler, one state owner and one source of truth for parameters. An extension adapts to these owners and may not introduce a second registry or runtime. The owners form seven layers, for semantics, declaration, resolution, execution, transactions, scientific structure and evidence, with an orthogonal control plane for lifecycle changes. The layers E1 to E5 classify scientific comparisons, whereas the seven software layers realize the obligations that those comparisons test.

\begin{table}[t]
\centering
\small
\caption[Isomorphism between mathematical contracts and RELIC architectural owners]{\textbf{Isomorphism between mathematical contracts and RELIC architectural owners.} Each mathematical obligation has one owner in the code and one mechanism that enforces it at compile time or runtime.}
\label{tab:owners}
\begin{tabularx}{\linewidth}{@{}>{\raggedright\arraybackslash}p{0.19\linewidth}>{\raggedright\arraybackslash}p{0.33\linewidth}>{\raggedright\arraybackslash}X@{}}
\toprule
Mathematical object & Owner in RELIC & Enforcement \\
\midrule
Quantity identity $\tau$ & \texttt{relic.contracts} (\texttt{ContractRegistry}); \texttt{PortSpec} & five field port comparison in \texttt{compile\_graph}; only registered conversions are inserted \\
Functional map $\mathcal F$ & \texttt{core/abstractions.py}, bridge classes, \texttt{FunctorBase} & every process implements one of four abstract contracts and is a \texttt{torch.nn.Module} \\
Parameters $\theta\in\Theta$ and constants & \texttt{ParameterManager} call stack; \texttt{physical/constants.yaml} & no constants in kernels; value, bounds, units and type declared in structured parameter files; learnable bindings keep their identity \\
Structure $\mathcal S$ (Definition~\ref{def:structure}) as a plan & \texttt{GraphIR}, \texttt{validate\_graph}, \texttt{compile\_graph} & topological steps, rate groups, writer ownership, strict precheck and plan seal \\
Declared input section & \texttt{GraphExecutor} call governor & a kernel receives only the state keys bound by its \texttt{ExecutionStep} \\
Governed evaluation $\Psi_{\mathrm{gov}}$ & \texttt{RELICSimulator.step}; \texttt{StateManager.apply\_updates} & typed \texttt{UpdateItem} proposals verified and committed at every rate group call; a rejected step is restored \\
Executed map $\mathcal T_{\mathrm{exec}}$ across devices & \texttt{core/device\_policy.py} & single owner of device resolution; device preparation deferred during compilation \\
External operators & \texttt{relic\_extension.yaml}; \texttt{PluginHost} & manifest verification; versioned C interface with load, retire, drain and unload \\
\bottomrule
\end{tabularx}
\end{table}

\subsection{Four abstract scientific contracts, FunctorBase and centralized parameters}

Every scientific process must satisfy one of four abstract contracts, \texttt{PhysicalField}, \texttt{BiologicalProcess}, \texttt{TransportMechanism} or \texttt{ScaleCoupling}, through its canonical bridge class, whose class name appends \texttt{Functor} to that of the contract. The four bridges converge on \texttt{FunctorBase}, which is an abstract process component class and a \texttt{torch.nn.Module}~\cite{Paszke2019} and supplies one entry point each for parameter access, dependency declaration, scientific metadata and conservation checks. An implementation that runs but has no abstraction is refused, so scientific meaning cannot dissolve into glue code. The base class also separates the arithmetic kernel from the governed call. The kernel computes rates and proposals from the inputs it is given and has no side effects; the governed call wraps it with input verification, a local conservation hook and output checks, and partitions what the kernel returns into state proposals, diagnostics and intents for external operations. This is the software form of $\Psi_{\mathrm{gov}}=\mathrm{post}\circ\Psi_{\mathrm{ker}}\circ\mathrm{pre}$ and of the three output spaces of Definition~\ref{def:function}; a direct call from one process to another is not part of the interface.

Parameters follow a call stack of four layers. Scientific modules call \texttt{get\_parameter\_value}, other code calls the injected manager, both resolve through the one \texttt{ParameterManager}, and the manager reads a versioned declarative database in which every entry declares value, bounds, units and data type with its literature source. Physical constants live in a single file. A learnable parameter is bound once as a tensor whose identity, gradient flag and device are retained when its value is updated, so the processes that hold it and an external optimizer share one object. Calibration, sensitivity analysis and gradient training therefore act on the same parameters without changes to scientific code, which is the property that differentiable Earth system modeling requires~\cite{Shen2023,Gelbrecht2023,Aboelyazeed2023,Feng2022}.

\subsection{Separation of physical, declared and execution graphs}

Three graphs are easily confused in coupled modeling (Fig.~\ref{fig:four}b). The physical exchange graph carries stocks and paired fluxes; it may contain feedback cycles ($\beta_1\ge1$), and conservation is verified on its ledgers through $\mathbf 1^\top\mathbf B=\mathbf 0$. The declared graph, a \texttt{GraphIR}, lists process nodes with typed ports, links, state bindings and parameter bindings. The execution plan is an acyclic order of steps within one outer time step. A water feedback between plant and soil is a cycle in the first graph and a sequence of reads and commits at stated time levels in the third, and the incidence structure of the execution plan is never a proof of physical conservation (Proposition~\ref{prop:static}). The distinction has measurable consequences: in a composition at ten stations whose demand link was left undeclared, the soil store read each demand one step late and 76\% of hourly soil water values changed, although every budget closed.

The compiler \texttt{compile\_graph} turns the declaration into a sealed plan without advancing any state. It verifies the declared graph, the registered implementations, ports, dependencies and writer ownership; lowers registered unit adapters onto edges; produces a topological list of \texttt{ExecutionStep} records in fast, medium and slow rate groups; audits structure; and then runs a strict compilation precheck, in which all processes are registered into a separate simulator that has no state manager and never steps. The precheck retains every class, bridge, port, writer, dependency and transaction check while deferring device preparation to the actual execution. This detail protects differentiation: moving a borrowed parameter tensor to another device during a compile check can sever it from a live \texttt{autograd} graph, and with deferral the association is preserved. A plan compiled without the precheck is refused by the executor. Compilation also reports, for every node, the inputs it binds and the optional inputs it leaves unbound together with the graph members that hold those names, so that an intended coupling which was never declared is visible before any step is taken.

\subsection{Declared input projection, conservation checked commits, bounded checkpoints and plugins}

\texttt{GraphExecutor} verifies the plan seal and precheck, builds the planned processes and binds a call governor. Inside the governor each kernel receives only the state keys that its step binds. An input that the step does not declare is absent for the kernel even when another node or a forcing field holds that name. What a process reads is therefore a subset of what its step declares, which is the premise of the ancestry invariance tested in Section~\ref{sec:verification}: if no declared ancestor of a process changes, its outputs cannot change.

\begin{figure}[tp]
\centering
\figfile{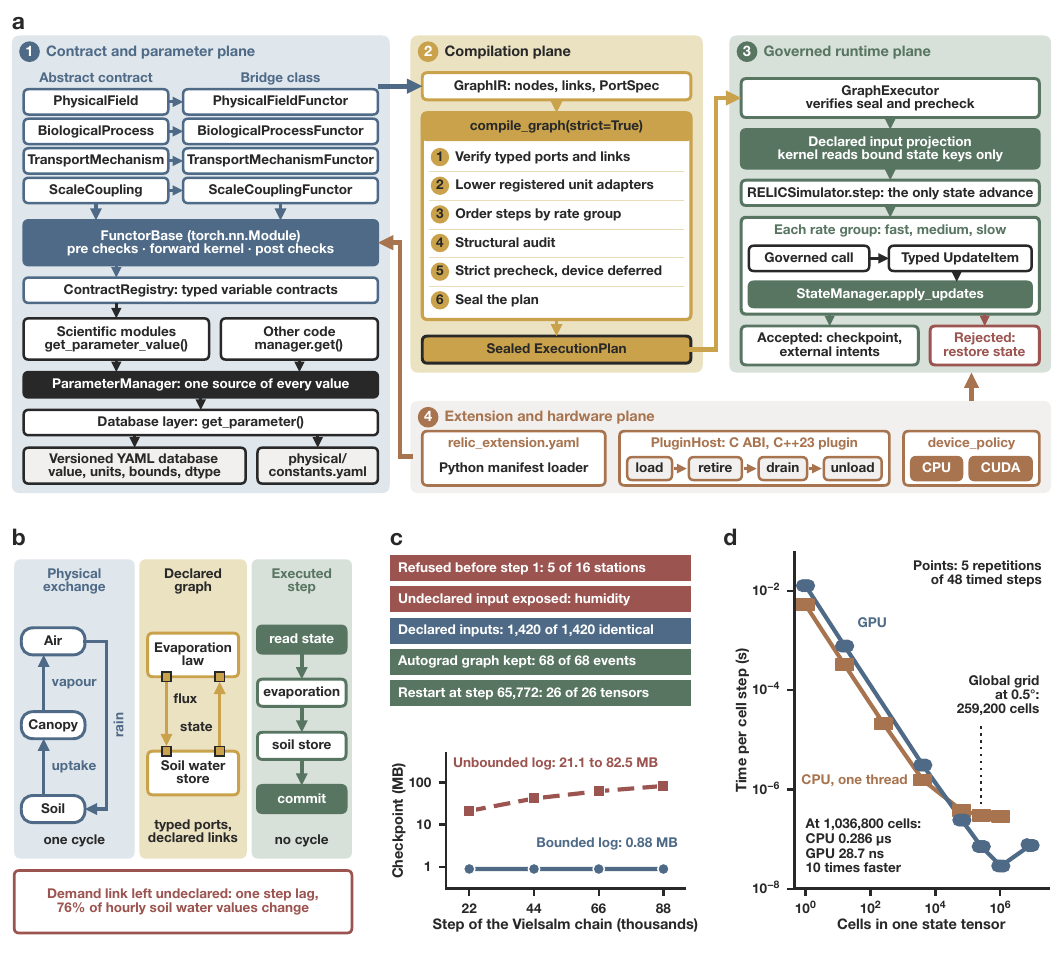}
\caption[Contract governance and differentiable graph compilation in RELIC]{\textbf{Contract governance and differentiable graph compilation in RELIC.} \textbf{a}, The four planes of the architecture: abstract scientific contracts, bridge classes, \texttt{FunctorBase} and the parameter call stack; the six stages of \texttt{compile\_graph} from \texttt{GraphIR} to the sealed plan; the governed runtime with declared input projection, typed proposals and commits at every rate group; and the extension and device plane. \textbf{b}, One water feedback between plant and soil as physical exchange graph with a cycle, as declared graph with typed ports and as acyclic executed step, with the measured effect of an undeclared demand link at ten stations. \textbf{c}, Governance outcomes in real workloads: refusals before the first step, an undeclared humidity input exposed by a handwritten loop, bitwise agreement of 1,420 series once only declared inputs are passed, preservation of the gradient graph in 68 of 68 probe events, a bitwise restart of 26 state tensors, and checkpoint size along the 87,696 steps of the Vielsalm chain with unbounded and bounded provenance. \textbf{d}, Time per cell and step of the governed step of four processes against the number of cells in one state tensor, on one thread of an AMD EPYC 9B45 processor and on an NVIDIA RTX PRO 6000 Blackwell graphics processor; points are five repetitions of 48 timed steps. Panels a and b are schematics. Datasets and benchmark configurations are summarized in Table~\ref{tab:provenance}.}
\label{fig:four}
\end{figure}

\texttt{RELICSimulator.step} is the only advance of scientific state. Each process returns typed proposals, diagnostics and intents. The proposals of a rate group are verified and committed by \texttt{StateManager.apply\_updates} when that group executes, with each symmetric substep of a Strang schedule as its own call, and there is no second commit at the end of the step. The accepted step is an outer transaction over state, numerical history, provenance, supported random streams and the meteorological forcing time index; a rejected step restores all of them. Checkpoints and deliveries to the outside world happen only after acceptance and carry their own receipts. Writer provenance is retained in a bounded window per variable, which keeps checkpoints at constant size while replay stays bitwise exact (Fig.~\ref{fig:four}c).

Changes to a running composition are governed in the same spirit. Registering, replacing or retiring a process publishes a new immutable view of the providers under a new epoch; a step reads through exactly one such view, and a transition is prepared before anything is mutated, with a receipt that records completion or compensation. Accepted scientific history, by contrast, is not reversible locally and can be revisited only through checkpoint and replay.

Device placement has one owner, \texttt{device\_policy}, which resolves processors and graphics devices and records completion fences, so the same governed step runs on batched tensors of one to several million cells (Fig.~\ref{fig:four}d). External code enters through two gates. Python processes are discovered through a manifest (\texttt{relic\_extension.yaml}). Native kernels, for instance C++23 or legacy Fortran behind a C interface, are hosted by \texttt{PluginHost}, which verifies the interface version at load and releases a library only after it has been retired and calls in flight have drained. Both gates adapt processes into the same lifecycle; neither adds a scheduler. Compared with couplers that exchange fields between components, the unit of governance here is the individual process and its typed ports, and compared with differentiable models assembled as scripts, conservation, state ownership and gradients are properties of the architecture and do not depend on the discipline of each author.

\section{Multiscale verification across flux stations and global land grids}
\label{sec:verification}

Empirical evaluation spans site networks, independent flux tower splits and global 0.5 degree land grids, with each benchmark retaining its declared denominator across layers E1 to E5 (Table~\ref{tab:verification}; Appendix~\ref{app:provenance}).

\subsection{Station network governance, interval enclosures and tensor scaling}
\label{sec:stations}

\paragraph{One composition at sixteen stations.} One declared plan of four processes (radiation, isoprene emission~\cite{Guenther2012}, Priestley--Taylor evapotranspiration~\cite{PriestleyTaylor1972} and a soil water bucket) was compiled with the inputs and parameters of 16 eddy covariance stations from tropical rainforest to boreal, semiarid and suburban sites. All 16 completed their driving windows, 288,912 steps in total (201,216 hourly and 87,696 at 30~min intervals). Before any step ran, the architecture refused 5 of the 16 stations, four wet stations whose initial soil water lay outside the bounds of the bucket and one station without a required leaf area input, each refusal naming the parameter or input; the stations ran once admissible soil parameters and a leaf area series were supplied. In the continuous chain across five years at Vielsalm the largest water residual of a single step was $1.50\times10^{-5}$~kg~m$^{-2}$, a restart from the checkpoint at step 65,772 reproduced all 26 state tensors bitwise, and the bounded provenance window kept the checkpoint at 0.88~MB while an unbounded ledger grew from 21.1 to 82.5~MB (Fig.~\ref{fig:four}c).

\begin{figure}[tp]
\centering
\figfile{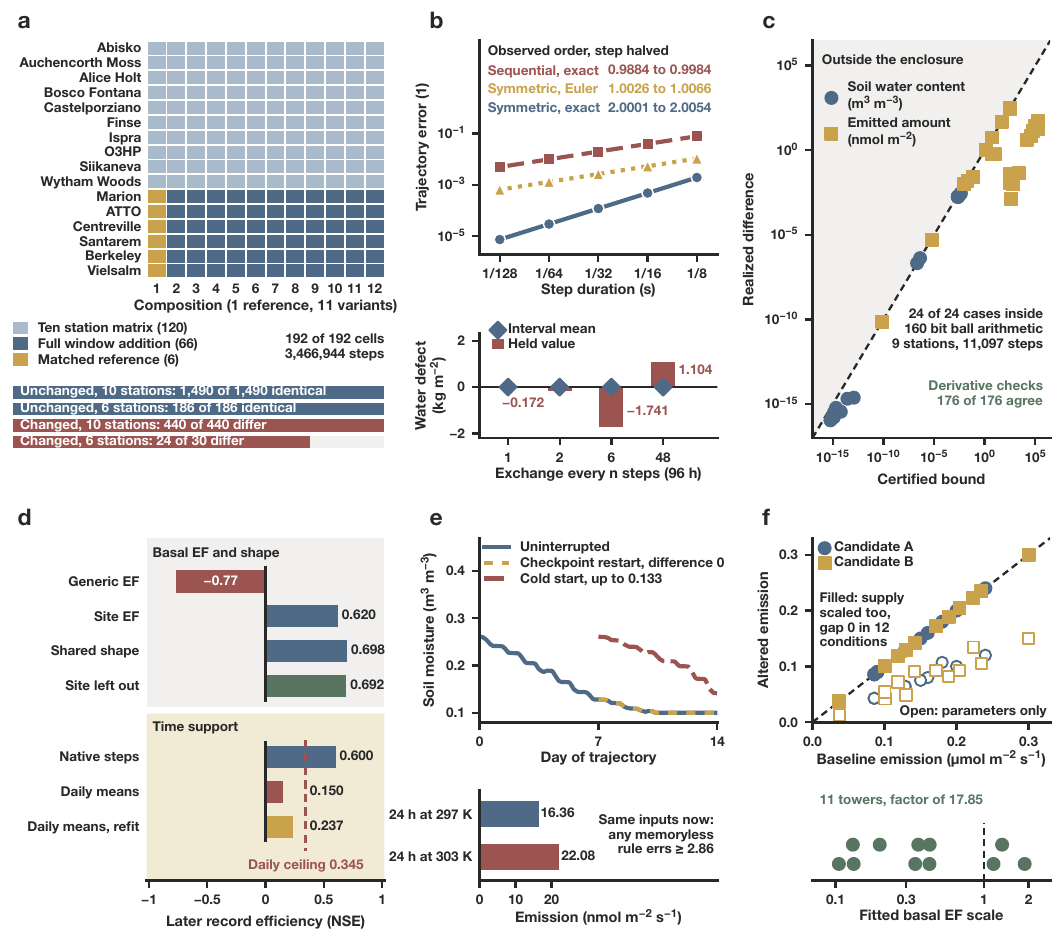}
\caption[Station network verification of ancestry invariance, interval bounds and identifiability]{\textbf{Station network verification of ancestry invariance, interval bounds and identifiability.} \textbf{a}, The 192 cells of 16 stations and 12 compositions (3,466,944 steps), colored by evaluation subset, with the counts of processes that stayed bitwise identical under unchanged ancestry and that differed under changed ancestry. Compositions: 1 reference, 2 Penman--Monteith, 3 photosynthesis added, 4 mechanistic emission, 5 both replacements, 6 lagged evapotranspiration, 7 symmetric schedule, 8 snow pack added, 9 thermal stress on photosynthesis, 10 water to carbon coupling, 11 soil water feedback on emission, 12 soil water and thermal stress on emission. \textbf{b}, Trajectory error against step duration for three splitting schedules with their observed orders, and the water transfer defect over 96 hours when held values or interval means are exchanged every $n$ steps. \textbf{c}, Realized differences from the discrete reference against their certified bounds for 24 enclosure cases, and the derivative comparisons. \textbf{d}, Nash--Sutcliffe efficiency (NSE) on 7,012 later records at 11 towers for emission factor (capacity) and shape treatments and, with an independently prepared driver set, for native and daily temporal support with the ceiling of any prediction that is constant within a day. \textbf{e}, Soil moisture of a stand of 16 patches after a checkpoint restart and after a cold start from the initial state (area mean), and emission for two reachable temperature histories under identical current inputs. \textbf{f}, Altered against baseline emission for two mechanistic candidates over 12 designed conditions, with and without scaling of the unobserved supply, and the fitted basal emission factor (capacity) multipliers of the 11 towers. Datasets and benchmark configurations are summarized in Table~\ref{tab:provenance}.}
\label{fig:five}
\end{figure}

\paragraph{Ancestry predicts invariance.} Twelve compositions, declared as lists of existing process classes and links, give 192 station and composition cells and 3,466,944 steps (Fig.~\ref{fig:five}a). Whenever a process kept its declaration and all its declared ancestors, its outputs stayed bitwise identical to the matched reference: 1,490 of 1,490 cases (13,470 port series) at ten stations and 186 of 186 cases (1,626 port series) at the other six, while all 440 processes with changed ancestry at the ten stations differed. A handwritten loop over the same kernels exposed an undeclared read of relative humidity by the Penman--Monteith process~\cite{Monteith1965}; with declared inputs only, loop and executor agreed bitwise in 1,420 of 1,420 series, the governed step taking 2.4 times as long. Bitwise statements hold within one processor family: reruns on AMD EPYC processors differed from an Intel Xeon reference in 12 of 33 port series; across such changes of hardware, statistical tests on ensembles take the place of bitwise comparison~\cite{ZemanSchar2022}.

\paragraph{Splitting order, holds, enclosures and derivatives.} A symmetric schedule with exact subflows converged at observed orders of 2.0001 to 2.0054 and the sequential schedule at 0.9884 to 0.9984, as Proposition~\ref{prop:static} requires. At one station, exchanging held values between rates every 2, 6 and 48 steps over 96 hours broke the physical water budget by $-0.172$, $-1.741$ and $+1.104$~kg~m$^{-2}$, whereas exchanging interval means kept the transfer closed; the failed variants are retained as negative results (Fig.~\ref{fig:five}b). For a declared discrete reference, exact rational arithmetic for the water map and outward 160 bit ball arithmetic for emission~\cite{Johansson2017Arb} enclosed the executed trajectories in 24 of 24 cases at nine stations (11,097 steps) at every saved interval; at Auchencorth Moss the largest difference in water content was 0.00282 within a bound of 0.00429~m$^3$~m$^{-3}$, and at Abisko the final difference in cumulative emission was 1.082 within 1.235~nmol~m$^{-2}$ (Fig.~\ref{fig:five}c). Enclosure widths differ by orders of magnitude between sites and condition their scientific use. In a state feedback experiment at nine stations, all 176 comparisons of automatic parameter derivatives with finite differences, chain reconstructions and conditional restarts passed; derivatives with respect to initial water passed 87 of 88 comparisons, the exception being a nonsmooth boundary with distinct directional derivatives~\cite{BoltePauwels2021}. With device preparation deferred during compilation, the original \texttt{autograd} graph survived all 68 probe events.

\paragraph{Scaling.} The same governed step was timed on state tensors of 1 to 1,036,800 cells on one processor thread and of up to 8,294,400 cells on a graphics processing unit (GPU; Fig.~\ref{fig:four}d). Replicating one station cell across each tensor isolates the cost of contract governance and batched kernel dispatch under uniform arithmetic load. At 1,036,800 cells, matching a global 0.25 degree grid, each cell step took 0.286~$\mu$s on the processor and 28.7~ns on the GPU; at 8,294,400 cells, GPU step time rose to 74.6~ns as peak allocated device memory reached 27.01~GB and spilled beyond the 128~MB level-2 cache into main memory bandwidth.

\subsection{Independent tower evaluation, the daily projection ceiling and local identifiability}
\label{sec:towers}

\paragraph{Shape transfers, emission factor stays local.} At 11 towers with resolved flux units (14,123 paired records, 7,111 for training and 7,012 later records for evaluation), the light, temperature and $\mathrm{CO_2}$ environmental activity response~\cite{Guenther2012} with its generic basal emission factor (emission capacity) failed on the evaluation records, with an NSE~\cite{NashSutcliffe1970} of $-0.77$. One basal emission factor scalar per site raised it to 0.620, three shared shape parameters to 0.698, and leaving the evaluated site out of shape fitting kept 0.692, or 92.6\% of the pooled gain (Fig.~\ref{fig:five}d). The fitted site emission factor (capacity) multipliers span 0.1055 to 1.8827, a factor of 17.85 partitioned into two clusters (0.11 to 0.43 at 8 towers and 1.16 to 1.88 at 3 towers). The short term response shape transfers between sites, whereas the absolute basal emission factor is local, which is the empirical signature of the equifinality orbit in Theorem~\ref{thm:equivalence}.

\paragraph{Temporal support and its ceiling.} Degrading the same evaluation to daily mean forcing, with a second, independently prepared set of drivers, lowered the NSE from 0.600 to 0.150, and to 0.237 after emission factors were refitted at daily support. For the observed records, the least squares projection onto arbitrary constants per site and day gives an exact ceiling of $\mathrm{NSE}\le0.345$ for every prediction that is constant within a day; this is Proposition~\ref{prop:three} in time. History matters in the same way. Two reachable 24~h temperature histories with identical current drivers give emissions of 16.36 and 22.08~nmol~m$^{-2}$~s$^{-1}$, so any predictor without that memory errs by at least 2.86~nmol~m$^{-2}$~s$^{-1}$ on one of them, and in a stand of 16 patches a cold start from the initial state instead of the full checkpoint shifted root zone soil moisture by up to 0.153~m$^3$~m$^{-3}$ in an individual patch (0.133~m$^3$~m$^{-3}$ in the stand area mean; Fig.~\ref{fig:five}e).

\paragraph{Compensation and rank.} For two mechanistic emission models coupled to photosynthesis~\cite{Farquhar1980,Morfopoulos2014}, scaling allocation parameters while scaling the unobserved electron or carbon supply inversely reproduced all 12 designed conditions with a largest difference of exactly 0.0 (Theorem~\ref{thm:equivalence}, part 3; Fig.~\ref{fig:five}f). On 1,723 records at Vielsalm~\cite{Dumont2026}, the scaled local Jacobian of emission with respect to three parameters of the coupling between stomatal ozone uptake and emission had rank 1 when only the net flux was observed. Adding the modeled leaf uptake raised it to 2 (singular values 113.23, 25.75 and 0), and only a further diagnostic of antioxidant consumption, which no tower yet measures, gave rank 3 (116.41, 48.81 and 19.90; condition number 5.849), as Proposition~\ref{prop:rank} predicts (Fig.~\ref{fig:two}h).

\subsection{Global 0.5 degree execution, satellite consilience and the commutation discrepancy}
\label{sec:global}

\begin{figure}[tp]
\centering
\figfile{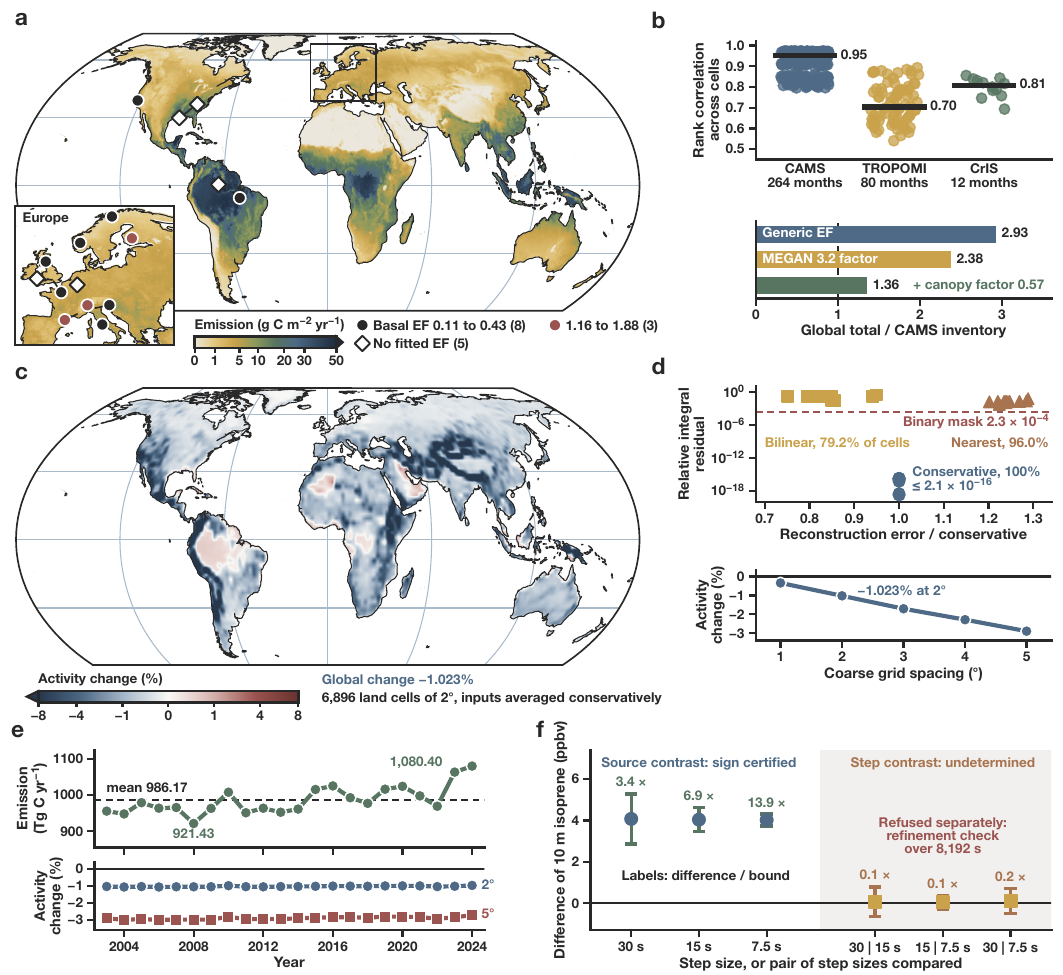}
\caption[Global 0.5 degree execution, conservative remapping and the discrepancy audit across scales]{\textbf{Global 0.5 degree execution, conservative remapping and the discrepancy audit across scales.} \textbf{a}, Mean annual isoprene emission of the governed baseline for 2003 to 2024 on its native 0.5 degree grid, the sum of 264 monthly maps over 61,951 cells with leaf area in at least one month, with the 16 audited flux stations: circles mark the 11 sites with a fitted basal emission factor (capacity) multiplier in two clusters (0.11 to 0.43 at 8 sites and 1.16 to 1.88 at 3 sites), and diamonds mark the 5 sites without fitted emission factor. \textbf{b}, Rank correlation across cells between the baseline and three independent products for every month, with medians, and the ratio of global totals to the CAMS-GLOB-BIO v4.0 inventory under three emission factor treatments. \textbf{c}, Change of isoprene activity over 264 months when daily temperature and shortwave radiation are averaged from 0.5 to 2 degrees, the commutation residual $e_F$, on 6,896 land cells. \textbf{d}, Relative integral residual against reconstruction error for 12 fields and three remapping methods in February 2023, and the global activity change against grid spacing with the central 95\% of 20,000 resamples of whole years. \textbf{e}, Annual global totals and the activity change at 2 and 5 degrees in each year. \textbf{f}, Differences of 10~m isoprene with their certified bounds for a source contrast and for contrasts between step sizes; labels give difference over bound. Both maps use the Equal Earth projection and are clipped to land; \textbf{a} shows the native cells without interpolation and \textbf{c} uses bilinear shading between the centers of its 2 degree cells. Datasets and benchmark configurations are summarized in Table~\ref{tab:provenance}.}
\label{fig:six}
\end{figure}

\paragraph{A governed global baseline.} Out of 259,200 cells on a global 0.5 degree grid and 95,491 land cells in the European Centre for Medium-Range Weather Forecasts Reanalysis version 5 Land (ERA5-Land) mask~\cite{MunozSabater2021}, the same governed operators ran on the 61,951 vegetated land cells carrying leaf area in at least one month (57,056 in every month) from 2003 to 2024 (264 months, 8,036 days). With a generic basal emission factor (capacity), the baseline emits 986.17~Tg~C per year on average, between 921.43~Tg~C in 2008 and 1,080.40~Tg~C in 2024, with maxima over Amazonia, the Congo basin, the Maritime Continent and the humid subtropics (Fig.~\ref{fig:six}a,e). Its spatial pattern agrees with independent products. The median rank correlation across cells is 0.70 with formaldehyde columns of the Tropospheric Monitoring Instrument (TROPOMI, 80 months)~\cite{DeSmedt2021,DeSmedt2025HCHO}, 0.81 with the isoprene climatology of the Cross-track Infrared Sounder (CrIS, 12 calendar months, 0.69 to 0.86)~\cite{Wells2020,Wells2022,WellsMillet2022CrIS} and 0.95 with the Copernicus Atmosphere Monitoring Service Global Biogenic Emissions (CAMS-GLOB-BIO) v4.0 inventory (264 months)~\cite{Sindelarova2022}. The global total, however, is 2.93 times that inventory, 2.38 times with the emission factors of the Model of Emissions of Gases and Aerosols from Nature version 3.2~\cite{Guenther2024MEGAN32,Bettineschi2026} and 1.36 times with a canopy coefficient of 0.57 in addition (Fig.~\ref{fig:six}b), and top down inversions of satellite formaldehyde likewise place totals below the generic baseline~\cite{Bauwens2016OMI,Sfendla2026}. This scale separation illustrates Theorem~\ref{thm:equivalence} at the planetary scale: forcing and response shape constrain the relative pattern in space and time, while the absolute level is a free factor until emission factor or canopy observations anchor it.

\paragraph{Conserved inputs, displaced response.} For 12 ERA5-Land fields on 95,491 land cells in February 2023, conservative area remapping~\cite{Jones1999,UllrichTaylor2015} from 0.5 to 2 degrees with fractional area weights~\cite{Taylor2024} kept the relative integral defect $d_I$ at or below $2.1\times10^{-16}$ and returned every cell; binary land masks left $2.3\times10^{-4}$ in a separate test. Bilinear interpolation gave the smaller reconstruction error $e_R$ on common cells (1.423 against 1.790~K for air temperature), returned only 79.2\% of the cells and did not conserve the integrals (Fig.~\ref{fig:six}d). Conservation of inputs to machine precision did not conserve the response. Averaging daily mean temperature and shortwave radiation conservatively from 0.5 degrees to coarser grids changed the integrated isoprene activity over all 264 months by $-0.341\%$, $-1.023\%$, $-1.707\%$, $-2.285\%$ and $-2.893\%$ at 1 to 5 degrees, the 2 degree change lying between $-1.033\%$ and $-1.012\%$ in 20,000 resamples of whole years, and the loss recurs in each of the 22 years (Fig.~\ref{fig:six}d,e). The map of this commutation residual $e_F$ on 6,896 land cells of 2 degrees is negative in 88.4\% of them, most strongly along the Andes, the margins of the Tibetan Plateau, the East African highlands and steep coasts, and weakly positive over parts of the Amazon lowlands and the Sahara (Fig.~\ref{fig:six}c); it is largest where fine resolution gradients of the drivers meet the curvature of the response~\cite{Jensen1906,rastetter1992}. The three quantities $d_I$, $e_R$ and $e_F$ thus separate on real global fields as Proposition~\ref{prop:three} states, and opposing regional signs show how aggregate cancellation can conceal local discrepancies (Remark~\ref{rem:inverse}).

\paragraph{A certified attribution.} In a column of six atmospheric layers, the mean isoprene mixing ratio at 10~m responded to a source increased by 50\% with differences of 4.07, 4.04 and 4.02~ppbv at steps of 30, 15 and 7.5~s, against a posteriori residual bounds~\cite{Estep1995Posteriori,Weyl2026} of 1.21, 0.58 and 0.29~ppbv. The response exceeds its certified envelope by factors of 3.4 to 13.9, so the criterion of Proposition~\ref{prop:attribution} holds at every step size and the sign of the source effect is certified (Fig.~\ref{fig:six}f). Differences between step sizes for one source, 0.04 to 0.11~ppbv, lie inside bounds of 0.35 to 0.72~ppbv, so the criterion is silent about them, as it should be. The record keeps its refusals as well: a separate refinement qualification over 8,192~s failed its check and was not used, and two temperature fields with one coarse state force any prediction from that coarse state to err by at least $1.458\times10^{-5}$~kg of tracer on one of them. A residual acquires the standing to speak about mechanism only when every preceding layer in the declared pipeline carries such certified bounds (Remark~\ref{rem:complete}).

\begin{table}[t]
\centering
\small
\caption[Multiscale verification summary across analytical, station and global benchmarks]{\textbf{Multiscale verification summary across analytical, station and global benchmarks.} Experiments, denominators, main results and the boundary of inference for each layer.}
\label{tab:verification}
\begin{tabularx}{\linewidth}{@{}l>{\raggedright\arraybackslash}p{0.27\linewidth}>{\raggedright\arraybackslash}p{0.33\linewidth}>{\raggedright\arraybackslash}X@{}}
\toprule
Layer & Experiment and denominator & Result & Boundary of inference \\
\midrule
E1 & 16 stations, 288,912 steps; 192 cells, 3,466,944 steps & 1,490 of 1,490 and 186 of 186 processes with unchanged ancestry bitwise identical; refusals before the first step & one processor family, batch shape and device \\
E2 & splitting test; 24 enclosure cases; 176 derivative comparisons & orders 2.0001 to 2.0054 and 0.9884 to 0.9984; 24 of 24 enclosed; 176 of 176 passed & declared discrete map and common invariant domain \\
E3 & 12 fields on 95,491 land cells; 264 months of global activity & $d_I\le2.1\times10^{-16}$; activity change $-1.023\%$ at 2 degrees and $-2.893\%$ at 5 degrees & daily mean drivers coarsened from 0.5 degrees \\
E4 & 7,012 later records at 11 towers; column of six layers & daily ceiling $\mathrm{NSE}\le0.345$; response 4.02 to 4.07 ppbv above bounds of 0.29 to 1.21 ppbv & archived clock mapping and footprints; declared column configuration \\
E5 & 11 towers; 12 designed conditions; 1,723 records; global grid over 264 months & NSE $-0.77$, 0.620, 0.698, 0.692; exact compensation; rank 1, 2, 3; median rank correlations 0.70 to 0.95 with level ratio 2.93 & emission factor (capacity) and supply share one observational class under flux data alone \\
\bottomrule
\end{tabularx}
\end{table}

\section{Synthesis and disciplinary horizon for verifiable Earth system science}
\label{sec:discussion}

\paragraph{One foundation for theory and software.} The separation of model conservation structure from process parameterizations, the five layer calculus and the governed graph architecture are three views of one object that answer the three questions of Section~\ref{sec:intro}. Together they give ecosystem and Earth system modeling a joint foundation of the kind that type systems give to programming languages and a posteriori analysis gives to numerical differential equations. Execution drift is locked out by compile time contracts and ancestry invariance (E1). Discretization and splitting across rates are quantified by gains and interval enclosures (E2). Aggregation across scales is resolved into the three mutually independent quantities $d_I$, $e_R$ and $e_F$ (E3). A candidate mechanism is refuted on the E5 axis once its residual crosses the envelope $\Gamma^M$ bounded on the E4 evaluation axis, whereas empirical support for one mechanism over a rival requires their separation to exceed $\Gamma^M+\Gamma^{M'}$ while the surviving candidate stays inside its envelope and satisfies the rank condition of Theorem~\ref{thm:equivalence} (Proposition~\ref{prop:attribution}). Verification, evaluation and confirmation, long distinguished in principle for models of open natural systems~\cite{oreskes1994}, thereby become separate computable obligations.

\paragraph{Consequences for ecophysiology and flux networks.} Adding mechanistic equations and refitting one scalar per site does not by itself improve extrapolation: the emission factor (capacity) refit at Vielsalm absorbed an unconstrained stress treatment, and the evaluation across 11 towers transferred short term response shape while local basal emission factors stayed local. The null space of the Jacobian and the multiplicative orbit state precisely what flux data cannot resolve, and they convert into a list of measurements. Leaf gas exchange and chlorophyll fluorescence anchor the supply that fluxes leave free; antioxidant metabolism supplies the direction that raised the local rank from 2 to 3; collocated gradients of nitrogen oxides and ozone at high sensitivity constrain the chemical side of the exchange. Network design~\cite{Heiskanen2022,Gettelman2022} can therefore be guided by rank and envelope calculations made before instruments are deployed, and a tower record can be reported together with the equivalence classes that it leaves open.

\paragraph{Consequences for land models and hybrid modeling.} For land surface and Earth system models~\cite{Lawrence2019,Bonan2024}, two distinctions carry practical weight. Conservation of input integrals does not bound the commutation error of a nonlinear response, which displaced global activity by 1\% at 2 degrees in Section~\ref{sec:global} while inputs were conserved to machine precision; and the cyclic physical graph is a different object from the acyclic execution order, so a missing declared link changes time levels without breaking any budget. For hybrid models that join physical processes with learned components~\cite{Reichstein2019,Kraft2022,Kochkov2024,Eyring2024Frontiers}, RELIC offers a safe substrate. A neural surrogate that implements one of the four bridge interfaces and declares the five tuple identity of its ports in the parameter and contract databases is subject to the same port proof, the same conservation checked state protection and the same gradient path as a mechanistic kernel, so a learned component can neither break a conservation ledger silently nor contaminate a branch that does not declare it as an ancestor. Because gradients pass through the same governed step that produces forward results, calibration against observations, sensitivity analysis and training of surrogates all act on the model that is actually executed.

\paragraph{Boundaries of inference and outlook.} Every conclusion holds inside four boundaries. The \emph{object} boundary: a ledger closes for a declared control volume, and the theorems apply within the domain and control volume that the contract declares. The \emph{applicability} boundary: constitutive laws hold in regimes of temperature and light, and the analytical benchmark cases prove that a shortcut can fail while leaving its frequency in nature to data. The \emph{scale} boundary: every reported number carries a grid spacing, a time step, a support and, for bitwise statements, a processor family and device; moving any of these defines a new comparison section. The \emph{inference} boundary: a flux design leaves basal emission capacity and a free multiplicative supply in one observational class, satellite inversions are themselves model products, and the global baseline shown here is a conditional run with a generic basal emission factor. Certificates for full ecological feedback loops and coverage of the complete software lifecycle remain open research obligations. Two companion studies build on this foundation: one quantifies the asymmetric effects of spatial and temporal scale on land water and carbon exchange, and one examines terrestrial water balance and bidirectional coupling between forests and tropospheric ozone from suburban margins to the global atmosphere.

\appendix
\section{Mathematical proofs and exact counterexamples}
\label{app:proofs}

\subsection{Proof of Theorem 1}

\emph{Step 1: the chain is defined.} By assumption $\pi z_0=Rx$, $Rx\in D_c$ and $x\in D_f$, and the executed, declared, coarse, fine and reference maps are evaluated on their declared domains, so $W_0,\dots,W_4$ all lie in $X_c$ and $H$ is defined at each of them.

\emph{Step 2: adjacent cancellation.} In the vector space $Y_{\mathrm{obs}}$,
\[
H(W_0)-H(W_4)=\sum_{k=1}^{4}\bigl(H(W_{k-1})-H(W_k)\bigr)=B_{\Eone}+B_{\Etwo}+B_{\Ethree}+B_{\Efive},
\]
because every intermediate value appears once with each sign. Since $y_{\mathrm{sim}}=H(W_0)$ and $y_{\mathrm{obs}}=H(W_4)+\eta_{\mathrm{obs}}$, subtracting gives part 1. No property of $H$, of the maps or of the distribution of the terms entered the argument.

\emph{Step 3: gains.} For E1, $\norm{B_{\Eone}}=\norm{H(W_0)-H(W_1)}\le L_H\norm{W_0-W_1}=L_H\varepsilon_{\Eone}$. The E2 and E3 terms follow from the same Lipschitz bound of $H$. For E5,
\[
\norm{B_{\Efive}}\le L_H\norm{R\Phi_f^{M}(x)-R\Phi_f^\star(x)}\le L_HL_R\norm{\Phi_f^{M}(x)-\Phi_f^\star(x)}\le L_HL_R\,\varepsilon_{\Efive},
\]
where the last step uses that the supremum over $D_f$ dominates the value at $x$. The triangle inequality applied to part 1 gives the total bound.

\emph{Step 4: cross terms.} With $C_1,\ldots,C_5$ as in the theorem, including $C_5=-\eta_{\mathrm{obs}}$, expanding $(\sum_i C_i)^2$ gives $\sum_i C_i^2+2\sum_{i<j}C_iC_j$. In an inner product space, bilinearity gives the corresponding identity with squared norms and cross inner products. Two unit scalar terms and zero remaining terms give a squared total of 4 and a sum of squares of 2. \qed

\emph{Example for Remark~\ref{rem:order}.} Take $f(a,b)=a^2b$ and replace $(a,b)=(2,2)$ by $(1,1)$ in two steps. Replacing $a$ first gives the terms $f(2,2)-f(1,2)=6$ and $f(1,2)-f(1,1)=1$; replacing $b$ first gives $f(2,2)-f(2,1)=4$ and $f(2,1)-f(1,1)=3$. Both sums equal 7, and the individual terms differ.

\subsection{Proof of Theorem 2}

\emph{Part 1.} For maps on a common domain, $\mathcal H_\theta=\mathcal H_{\theta'}$ means equality at every point of $D_f$. Pointwise equality is reflexive, symmetric and transitive, so $\sim_{\mathcal M}$ is an equivalence relation, and the classes of an equivalence relation partition its carrier set. All classes are singletons exactly when no two distinct parameters share an observed map, which is injectivity of $\theta\mapsto\mathcal H_\theta$ on $\Theta$. If $\mathcal H_\theta=\mathcal H_{\theta'}$ on $D_f$, then the two maps agree at every design point, so $\mathcal G(\theta)=\mathcal G(\theta')$: each class of $\sim_{\mathcal M}$ lies inside a class of the finite design.

\emph{Part 2.} If $\mathcal G(\theta(\sigma))$ is constant along a differentiable curve $\sigma\mapsto\theta(\sigma)$, the chain rule gives $0=\tfrac{\mathrm d}{\mathrm d\sigma}\mathcal G(\theta(\sigma))|_{\sigma=0}=J(\theta(0))\,\theta'(0)$. If $J_0=J(\theta_0)$ has full column rank, $J_0^\top J_0$ is positive definite and $P=(J_0^\top J_0)^{-1}J_0^\top$ satisfies $PJ_0=I_p$. By continuity of $J$ there is a convex open ball $U\ni\theta_0$ with $\norm{PJ(\theta)-I_p}_2\le\tfrac12$ on $U$. For $u,v\in U$, integrating along $u+t(v-u)$ gives
\[
P[\mathcal G(v)-\mathcal G(u)]=(v-u)+\int_0^1[PJ(u+t(v-u))-I_p](v-u)\,\mathrm dt ,
\]
hence $\norm{P[\mathcal G(v)-\mathcal G(u)]}_2\ge\tfrac12\norm{v-u}_2>0$ for $u\ne v$. The map $\theta\mapsto\theta^3$ has $J(0)=0$ and is injective, so full rank is not necessary.

\emph{Part 3.} For $a>0$, $s/a$ is again a positive function and $aV>0$, and $(s(t)/a)\,(aV)\,g(C(t);\vartheta)=s(t)\,V\,g(C(t);\vartheta)$ at every time, so the two pairs give the same trajectory and lie in one class. If $s$ is fixed independently of $V$, $\vartheta$ is known, $s(t_1)\ne0$ and $g(C(t_1);\vartheta)\ne0$, dividing the observation equation at $t_1$ by $s(t_1)g(C(t_1);\vartheta)$ determines $V$. \qed

\subsection{Assumptions and proof of Proposition 9}

The proof uses only the identity of Theorem~\ref{thm:telescope}, part 1, the triangle inequality and the reverse triangle inequality $\norm{u-v}\ge\norm{u}-\norm{v}$.

\begin{enumerate}
\item[(A1)] \emph{Chain.} $X_f$ and $X_c$ are real normed spaces of finite dimension with declared domains $D_f\subseteq X_f$ and $D_c\subseteq X_c$, and $Y_{\mathrm{obs}}$ is a real normed space. For the candidate $M$ and the evaluated state $x\in D_f$ one has $Rx\in D_c$, a runtime state $z_0$ with $\pi z_0=Rx$, and the five states $W_0,\dots,W_4\in X_c$ of Section~\ref{sec:calculus}, with $W_3=R\Phi_f^{M}(x)$, are defined. The observation operator $H$ is defined at $W_0,\dots,W_4$. The simulated and observed outputs are $y^{M}_{\mathrm{sim}}=H(W_0)$ and $y_{\mathrm{obs}}=H(W_4)+\eta_{\mathrm{obs}}$ with $\eta_{\mathrm{obs}}\in Y_{\mathrm{obs}}$.
\item[(A2)] \emph{Gains.} There are certified constants $L_H>0$ and $L_R>0$ such that $\norm{H(u)-H(v)}\le L_H\norm{u-v}$ for $u,v\in\{W_3,W_4\}$ and $\norm{R(p)-R(q)}\le L_R\norm{p-q}$ for $p,q\in\{\Phi_f^{M}(x),\Phi_f^\star(x)\}$.
\item[(A3)] \emph{Envelope.} The numbers $b_1,b_2,b_3,\bar\eta\ge0$ satisfy $\norm{B_{\Eone}}\le b_1$, $\norm{B_{\Etwo}}\le b_2$, $\norm{B_{\Ethree}}\le b_3$ and $\norm{\eta_{\mathrm{obs}}}\le\bar\eta$, and $\Gamma^{M}=b_1+b_2+b_3+\bar\eta$.
\item[(A4)] \emph{Common observation.} Two candidates $M$ and $M'$ each satisfy (A1) to (A3) with their own chains, gains, bounds and envelopes, and with the same spaces, the same operators $R$ and $H$, the same declared domains, the same evaluated state $x$, the same reference process $\Phi_f^\star$ and the same observation error $\eta_{\mathrm{obs}}$. Hence $W^{M}_4=W^{M'}_4$ and both are compared with one $y_{\mathrm{obs}}$.
\item[(A5)] \emph{Reference candidate.} Candidate $M$ reproduces the reference at the evaluated state, $\Phi^{M}_{f}(x)=\Phi_f^\star(x)$.
\end{enumerate}

\emph{Part 1.} By (A1) and Theorem~\ref{thm:telescope}, part 1, $r^{M}=B_{\Eone}+B_{\Etwo}+B_{\Ethree}+B_{\Efive}-\eta_{\mathrm{obs}}$, so $B_{\Efive}=r^{M}-v$ with $v=B_{\Eone}+B_{\Etwo}+B_{\Ethree}-\eta_{\mathrm{obs}}$. The reverse triangle inequality, the triangle inequality and (A3) give
\[
\norm{B_{\Efive}}\ \ge\ \norm{r^{M}}-\norm{v}\ \ge\ \norm{r^{M}}-\bigl(\norm{B_{\Eone}}+\norm{B_{\Etwo}}+\norm{B_{\Ethree}}+\norm{\eta_{\mathrm{obs}}}\bigr)\ \ge\ \norm{r^{M}}-\Gamma^{M}.
\]
By (A2), $\norm{B_{\Efive}}\le L_H\norm{R\Phi_f^{M}(x)-R\Phi_f^\star(x)}\le L_HL_R\norm{\Phi_f^{M}(x)-\Phi_f^\star(x)}$. If $\norm{r^{M}}>\Gamma^{M}$, dividing by $L_HL_R>0$ gives the stated bound, which is positive; since $x\in D_f$, $\varepsilon_{\Efive}$ is at least the value at $x$ and is positive. When $\norm{r^{M}}\le\Gamma^{M}$ both cases occur. If $\Phi_f^{M}=\Phi_f^\star$ on $D_f$ and all other terms are zero, then $r^{M}=0$ and $\varepsilon_{\Efive}=0$. If instead $B_{\Efive}=-(B_{\Eone}+B_{\Etwo}+B_{\Ethree})\ne0$ and $\eta_{\mathrm{obs}}=0$, then $r^{M}=0$ while $\varepsilon_{\Efive}>0$ (Proposition~\ref{prop:cascade}).

\emph{Part 2.} By (A4) both residuals use one $y_{\mathrm{obs}}$, so $y^{M}_{\mathrm{sim}}-y^{M'}_{\mathrm{sim}}=r^{M}-r^{M'}$. Suppose part 1 applied to neither candidate, that is $\norm{r^{M}}\le\Gamma^{M}$ and $\norm{r^{M'}}\le\Gamma^{M'}$. Then $\Delta^{M,M'}\le\norm{r^{M}}+\norm{r^{M'}}\le\Gamma^{M}+\Gamma^{M'}$, which contradicts the criterion. Under (A5), $W^{M}_3=W_4$, so $B^{M}_{\Efive}=0$ and $\norm{r^{M}}\le\Gamma^{M}$; hence $\norm{r^{M'}}>\Gamma^{M'}$ and part 1 applies to $M'$. The condition is sufficient only: a pair that fails it may still differ in mechanism.

\emph{Part 3.} If $\theta\sim_{\mathcal M}\theta'$, then $H(R\Phi_{f,\theta}(x))=H(R\Phi_{f,\theta'}(x))$ because $x\in D_f$, so under (A4) the two chains share $H(W_3)$ and $W_4$, and $B^{\theta}_{\Efive}=B^{\theta'}_{\Efive}$. Adjacent cancellation over the first three replacements gives $y^{\theta}_{\mathrm{sim}}=H(W^{\theta}_3)+B^{\theta}_{\Eone}+B^{\theta}_{\Etwo}+B^{\theta}_{\Ethree}$, and likewise for $\theta'$, so
\[
\Delta^{\theta,\theta'}=\norm{\textstyle\sum_{k\le3}B^{\theta}_{\mathrm Ek}-\sum_{k\le3}B^{\theta'}_{\mathrm Ek}}\le b^{\theta}_1+b^{\theta}_2+b^{\theta}_3+b^{\theta'}_1+b^{\theta'}_2+b^{\theta'}_3\le\Gamma^{\theta}+\Gamma^{\theta'},
\]
and the criterion fails.

\emph{Part 4.} If part 1 applies to $\theta$, its proof gives $\norm{B^{\theta}_{\Efive}}\ge\norm{r^{\theta}}-\Gamma^{\theta}>0$. Because $\theta'$ uses the same $H$, $R$, $x$ and $\Phi_f^\star$, $B^{\theta'}_{\Efive}=\mathcal H_{\theta'}(x)-H(R\Phi_f^\star(x))=\mathcal H_{\theta}(x)-H(R\Phi_f^\star(x))=B^{\theta}_{\Efive}$, and (A2) for $\theta'$ gives $\norm{B^{\theta'}_{\Efive}}\le L'_HL'_R\norm{\Phi_{f,\theta'}(x)-\Phi_f^\star(x)}$. Dividing by $L'_HL'_R>0$ gives the stated bound. \qed

\emph{Why a positive tolerance does not transfer.} Take $X_f=X_c=\mathbb R^2$ with the maximum norm, $R=\mathrm{id}$, $H(u,v)=u+v/4$, the reference state $(0,0)$ and a zero envelope. A member $\theta$ with fine state $(0,4)$ and a member $\theta'$ with fine state $(1,0)$ both give the observed value 1, so their observed maps coincide at $x$. On the pair of states named in (A2), $\theta$ has $L_H=1/4$ and $L_R=1$, and part 1 certifies a discrepancy of at least $1/(1/4)=4$, its true value. For $\theta'$ the gain is $L'_H=1$, and part 4 transfers the bound 1, again its true value. With $\delta_5=2$ the first member is excluded while the second lies within tolerance. A positive tolerance therefore needs a bound for each member or a common gain bound over the class.

\section{Formal specification of RELIC compilation and runtime contracts}
\label{app:specs}

The rules below restate, in the notation of Sections~\ref{sec:theory} and~\ref{sec:architecture}, the contracts that the RELIC compiler and runtime enforce. They specify the software; the measured evidence that the implementation obeys them is in Section~\ref{sec:verification}.

\subsection{Port compatibility}

For an output port $p$ and an input port $q$ with identities $\tau_p,\tau_q$ (Definition~\ref{def:type}) and a registry $\mathcal K$ of conversions, each with a declared domain, the classification of Proposition~\ref{prop:static} is
\[
T(p,q)=
\begin{cases}
\mathrm{Exact} & \text{if } \tau_p=\tau_q \text{ in all five fields,}\\
\mathrm{Castable} & \text{if } \tau_p\ne\tau_q \text{ and some } \kappa\in\mathcal K \text{ maps } \tau_p \text{ to } \tau_q \text{ on its declared domain,}\\
\mathrm{Rejected} & \text{otherwise.}
\end{cases}
\]
A castable link is realized by inserting the registered adapter $\kappa$ as an explicit node on the edge. Equality of unit strings with any other field unequal is never $\mathrm{Exact}$, and no conversion is inferred from code.

\subsection{The sealed compilation pipeline}

\texttt{compile\_graph} maps a declared graph to a sealed plan in six stages and advances no state.
\begin{enumerate}
\item[(C1)] \emph{Verification.} The declared graph, the registered implementations, ports, dependencies, writer ownership and assertions are checked; every executable node must resolve to a process class behind one of the four bridges.
\item[(C2)] \emph{Adapter lowering.} Every castable link receives its registered adapter node; rejected links stop compilation.
\item[(C3)] \emph{Ordering.} A topological list of \texttt{ExecutionStep} records is produced. Each step carries its bound inputs (links or state keys), its rate group (fast, medium, slow or none; any other value is refused) and its parameter bindings.
\item[(C4)] \emph{Structural audit.} Declared structural assertions are evaluated, and the input projection of every node, including optional inputs that it does not bind, is disclosed in the compile report.
\item[(C5)] \emph{Strict precheck.} All planned processes are registered into a separate simulator that has no state manager and never steps. Class, bridge, port, writer, dependency, transaction and publication checks run with device preparation deferred, so borrowed parameter tensors keep their device and their association with an existing gradient graph.
\item[(C6)] \emph{Seal.} The plan is sealed together with the precheck certificate. The executor refuses a plan whose seal does not verify or whose precheck was skipped.
\end{enumerate}

\subsection{Runtime invariants}

Let step $i$ of a sealed plan bind the state keys $K_i$, and let $\Omega$ denote the shared state.
\begin{enumerate}
\item[(R1)] \emph{Single advance.} \texttt{RELICSimulator.step} is the only advance of scientific state, and each prepared run uses one fresh simulator.
\item[(R2)] \emph{Declared input projection.} The kernel of step $i$ receives $\Omega|_{K_i}$. An unbound optional input is absent even when another node or a forcing field holds its name, and bookkeeping keys are absent as well. State verification and the conservation hook of the governed call still read the whole state.
\item[(R3)] \emph{Order.} Rate groups run fast, then medium, then slow. Inside a group the order is topological over the edges of that group, with ties broken by registration index. Dependencies across groups and lagged dependencies are time level declarations and are not ordered by the graph.
\item[(R4)] \emph{Group commit.} The typed proposals of a group are lowered to \texttt{UpdateItem} records and committed by \texttt{StateManager.apply\_updates} when the group executes; every symmetric substep of a Strang schedule is its own call. No second commit follows at the end of the step.
\item[(R5)] \emph{Step transaction.} A savepoint covers state, runtime, process state, provenance, supported random streams and the meteorological forcing time index. A rejected step restores all of them. Checkpoints and external deliveries run only after acceptance, are retried and receipted, and are never rolled into the scientific step.
\item[(R6)] \emph{Bounded provenance.} After acceptance each variable keeps its most recent $M$ writer records (256 by default); the last writer is always retained, and a rejected step restores the previous window exactly.
\item[(R7)] \emph{Device ownership.} Device resolution, overrides and completion fences have one owner. A device transfer at registration waits for a fence on the source device.
\item[(R8)] \emph{Native lifecycle.} A native library is bound and its interface version verified at load; retirement refuses new calls while calls in flight continue; the library is released only when the count of calls in flight has drained to zero, and an expired drain leaves an observable receipt.
\end{enumerate}

\paragraph{Consequence tested in Section 6.} Under (R1) to (R4), for deterministic kernels on one processor family, batch shape and device, a process whose declaration, bound inputs and parameter bindings are unchanged between two plans, and all of whose declared ancestors are unchanged, receives bitwise equal inputs at every step and returns bitwise equal outputs. The 1,490 of 1,490 and 186 of 186 cases of Section~\ref{sec:stations} test this consequence; the hidden humidity read found there is what a violation of (R2) looks like in a handwritten loop.

\section{Empirical datasets and multiscale benchmark inventory}
\label{app:provenance}

Table~\ref{tab:provenance} summarizes the observational datasets, meteorological forcings, satellite retrievals, analytical counterexamples and benchmark configurations underlying Figures~\ref{fig:one} to~\ref{fig:six} and Tables~\ref{tab:layers} to~\ref{tab:verification}, together with the sample support or denominator of every comparison. Panels marked analytic or schematic carry closed form expressions or structural specifications without empirical observations.

\paragraph{Data, code and competing interests.} Eddy covariance flux records, ERA5-Land meteorological fields, satellite leaf area indices, TROPOMI formaldehyde columns, CrIS isoprene retrievals and the CAMS-GLOB-BIO v4.0 emission inventory are publicly available from their respective archives cited in the text; non-public raw subdaily station measurement records governed by site data-sharing agreements are available from the corresponding author upon reasonable request. The Recomposable Ecosystem Library for Integrated Coupling (RELIC, release v1.0.0), interactive tutorials, bilingual documentation (\url{https://ayinmostima.github.io/RELIC/}) and bilingual reference volumes (\emph{Methods of Ecological Analysis}) are publicly available at \url{https://github.com/AyinMostima/RELIC} under the MIT License. The authors declare no competing financial or non-financial interests related to the research, authorship or publication of this study.

{\footnotesize
\begin{longtable}{@{}>{\raggedright\arraybackslash}p{0.08\linewidth}>{\raggedright\arraybackslash}p{0.25\linewidth}>{\raggedright\arraybackslash}p{0.46\linewidth}>{\raggedright\arraybackslash}p{0.13\linewidth}@{}}
\caption[Empirical datasets and benchmark configurations of the figures and tables]{\textbf{Empirical datasets and benchmark configurations of the figures and tables.} Evaluated quantity, empirical dataset or benchmark configuration, and sample support for each figure panel and table.}\label{tab:provenance}\\
\toprule
Panel & Evaluated quantity & Empirical dataset or benchmark configuration & Sample support \\
\midrule
\endfirsthead
\toprule
Panel & Evaluated quantity & Empirical dataset or benchmark configuration & Sample support \\
\midrule
\endhead
1a to 1c & multisphere coupling, contracts and replacement chain & structural sextuple, five tuple port schema and replacement sequence $W_0$ to $W_4$ & schematic \\
1d & refitted emission factor and evaluation error at Vielsalm & chronological eddy covariance flux split at Vielsalm forest with and without stomatal ozone stress & 861 and 862 records \\
\midrule
2a to 2g & analytical benchmark cases of Propositions~\ref{prop:unit} to~\ref{prop:quant} & closed form algebraic, differential, splitting, aggregation and network counterexamples & analytic \\
2h & singular values and local Jacobian rank & Vielsalm forest eddy covariance and leaf ozone uptake observation sets & 1,723 records \\
\midrule
3a to 3f & geometry of Theorems~\ref{thm:telescope} and~\ref{thm:equivalence} and Proposition~\ref{prop:attribution} & closed form replacement chains, hyperbolic equivalence orbits and discrimination domains & analytic \\
\midrule
4a, 4b & governed graph planes and undeclared demand link & four plane architecture specification and plant to soil water coupling across eddy covariance sites & 10 stations \\
4c & admission refusals, input isolation, gradient probes and checkpoint window & full period station compilation, kernel isolation test, five year Vielsalm integration and autograd probes & 16 stations; 87,696 steps; 68 events \\
4d & execution time per cell and step & replicated four process tensor scaling on one processor thread and a graphics processor & 5 repetitions of 48 steps \\
\midrule
5a & station by composition matrix and ancestry invariance & 12 process compositions compiled across 16 eddy covariance stations from tropical to boreal sites & 192 cells; 3,466,944 steps \\
5b & splitting convergence orders and multirate hold defects & sequential and symmetric operator splitting sweep and 96 hour multirate water exchange test & one station, 96 hours \\
5c & certified trajectory enclosures and automatic derivatives & exact rational soil water arithmetic, 160 bit ball emission enclosures and state feedback tangent tests & 24 cases; 176 comparisons \\
5d & independent tower evaluation and daily projection ceiling & chronological flux split across eddy covariance towers under subdaily and daily forcing & 11 towers; 7,012 records \\
5e & state memory lower bound and stand checkpoint restart & reachable 24 hour thermal histories and 16 patch stand moisture integration & 2 histories; 16 patches \\
5f & hyperbolic supply compensation and site emission factor clusters & coupled photosynthesis and isoprene allocation experiments and fitted tower emission factors & 12 conditions; 11 sites \\
\midrule
6a, 6e & global 0.5 degree baseline emission and annual totals & governed global land simulation driven by ERA5-Land meteorology and satellite leaf area from 2003 to 2024 & 61,951 cells of 0.5 degrees; 264 months \\
6b & satellite formaldehyde, infrared isoprene and inventory comparison & TROPOMI formaldehyde columns, CrIS isoprene climatology and CAMS-GLOB-BIO v4.0 monthly inventory & 80, 12 and 264 months \\
6c, 6e & spatial commutation residual map and annual coarsening shift & conservative aggregation of daily temperature and shortwave radiation from 0.5 to 2 and 5 degrees & 6,896 cells; 264 months \\
6d & remapping conservation defect and multiresolution curve & conservative, bilinear and nearest-neighbor remapping of ERA5-Land fields in February 2023 and 22 year bootstrap & 12 fields; 95,491 cells \\
6f & certified 10 m mixing ratio contrast against residual bounds & six layer atmospheric diffusion and source column at 30, 15 and 7.5 s steps & 3 step sizes \\
\midrule
Table 1 & five layer definitions and analytical benchmark cases & theoretical framework and counterexamples of Sections~\ref{sec:theory} and~\ref{sec:limits} & analytic \\
Table 2 & terminology bridge across Earth system frameworks & correspondence to ESMF, NUOPC, BMI, SUMMA and Oreskes boundaries & specification \\
Table 3 & compile time and runtime contract owners & formal compilation and runtime specification of Appendix~\ref{app:specs} & specification \\
Table 4 & multiscale verification summary & analytical, station network and global benchmarks of Figures~\ref{fig:four} to~\ref{fig:six} & E1 to E5 \\
\bottomrule
\end{longtable}
}

\bibliographystyle{unsrtnat}
\bibliography{references}
\end{document}